\documentstyle[emulateapj,psfig]{article}

\received{}
\revised{}
\accepted{}

\cpright{}{}
\journalid{}{}
\articleid{}{}
\paperid{}

\cpright{}{}
\ccc{}

\lefthead{Alcock, C. \& the {\sc macho} collaboration}
\righthead{FREQUENCY ANALYSIS OF THE RR LYRAE STARS IN THE LMC}

\begin{document}

\title{THE MACHO PROJECT LMC VARIABLE STAR INVENTORY. IX. 
FREQUENCY ANALYSIS OF THE FIRST OVERTONE RR LYRAE STARS AND THE 
INDICATION FOR NONRADIAL PULSATIONS
}

\author{Alcock, C.$^1$, Allsman, R.$^2$, Alves, D. R.$^3$, 
Axelrod, T.$^2$, Becker, A.$^4$, Bennett, D.$^{12}$, Clement, C.$^{15}$, 
Cook, K.H.$^1$, Drake, A.,$^2$, Freeman, K.$^2$, Geha, M.$^1$, 
Griest, K.$^5$, Kov\'acs, G.$^{13}$, Kurtz, D.W.$^{14}$, Lehner, M.$^{11}$, 
Marshall, S.$^{1}$, Minniti, D.$^6$, Nelson, C.$^{1,7}$, Peterson, B.$^2$, 
Popowski, P.$^1$, Pratt, M.$^4$, Quinn, P.$^8$, Rodgers, A.$^{16}$, 
Rowe, J.$^{15}$, Stubbs, C.$^4$, Sutherland, W.$^9$, Tomaney, A.$^4$, 
Vandehei, T.$^5$ and Welch, D.L.$^{10}$}

\footnotetext[1]{Institute for Geophysics and Planetary Physics, 
Lawrence Livermore National Laboratory, Livermore, CA 94550, USA \\
 alcock, kcook, mgeha, stuart, cnelson, popowski1@llnl.gov}
\footnotetext[2]{Mount Stromlo and Siding Spring Observatory, 
Australian National University, Weston Creek, ACT 0200, Australia \\
 ajd, kcf, peterson, tsa@mso.anu.edu.au, robyn@mindful.anu.edu.au}
\footnotetext[3]{Space Telescope Science Institute, 
Baltimore, MD 21218, USA \\
 alves@stsci.edu}
\footnotetext[4]{Department of Astronomy, University of Washington, 
Seattle, WA 98195, USA \\
 austin, becker, stubbs@astro.washington.edu}
\footnotetext[5]{Department of Physics, University of California, 
San Diego, CA 92093, USA \\
 kgriest, tvandehei@ucsd.edu}
\footnotetext[6]{Department Astronomia, Universidad Catolica, Chile \\
 dante@astro.puc.cl}
\footnotetext[7]{Department of Physics, University of California, 
Berkeley, CA 94720, USA}
\footnotetext[8]{European Southern Observatory, Garching, D-85748, Germany \\
 pjq@eso.org}
\footnotetext[9]{Department of Physics, Oxford University, OX1 3RH, England \\
 w.sutherland@physics.ox.ac.uk}
\footnotetext[10]{Department of Physics and Astronomy, McMaster University, 
Hamilton, L82 4M1, Canada \\
 welch@physics.mcmaster.ca}
\footnotetext[11]{Department of Physics, University of Sheffield, 
S3 7RH, England}
\footnotetext[12]{Department of Physics, University of Notre Dame, 
South Bend, IN 46556, USA \\
 bennett@nd.edu}
\footnotetext[13]{Konkoly Observatory, P.O. Box 67, H-1525 Budapest, 
Hungary \\
 kovacs@konkoly.hu}
\footnotetext[14]{Department of Astronomy, University of Cape Town, 
Rondebosch 7701, South Africa \\
 dkurtz@ma.saao.ac.za}
\footnotetext[15]{Department of Astronomy, University of Toronto, 
Toronto, M5S 3H8, Canada \\
 cclement, rowe@astro.utoronto.ca}
\footnotetext[16]{deceased}

\begin{abstract}
More than 1300 variables classified provisionally as first overtone 
RR~Lyrae pulsators in the {\sc macho} variable star database of the Large 
Magellanic Cloud (LMC) have been subjected to standard frequency analysis. 
Based on the remnant power in the prewhitened spectra, we found 70\% 
of the total population to be monoperiodic. The remaining 30\% (411 stars) 
are classified as one of 9 types according to their frequency spectra. 
Several types of RR~Lyrae pulsational behavior are clearly identified 
here for the first time. Together with the earlier discovered 
double-mode (fundamental \& first overtone) variables this study 
increased the number of the known double-mode stars in the LMC to 181. 
During the total 6.5~yr time span of the data, 10\% of the stars show 
strong period changes. The size, and in general also the patterns of 
the period changes exclude simple evolutionary explanation. We also 
discovered two additional types of multifrequency pulsators with low 
occurrence rates of 2\% for each. In the first type there remains one 
closely spaced component after prewhitening by the main pulsation 
frequency. In the second type the number of remnant components is two, 
they are also closely spaced, and, in addition, they are symmetric in 
their frequency spacing relative to the central component. This latter 
type of variables is associated with their relatives among the 
fundamental pulsators, known as Blazhko variables. Their high 
frequency ($\approx 20\%$) among the fundamental mode variables 
versus the low occurrence rate of their first overtone counterparts 
makes it more difficult to explain Blazhko phenomenon by any theory 
depending mainly on the role of aspect angle or magnetic field. None 
of the current theoretical models are able to explain the observed 
close frequency components without invoking nonradial pulsation 
components in these stars.     
\end{abstract} 
%
%
\section{INTRODUCTION}

In this continuing series of papers dealing with the variable star 
data of the {\sc macho} project, we examine here the temporal behavior 
of a large sample of {\it first overtone RR~Lyrae} stars. Until very 
recently, short-periodic RR~Lyrae stars have been known to appear in 
two different flavors: (a) singly-periodic; (b) doubly-periodic 
(or double-mode). In this latter case the two periods are 
associated with the first two radial normal modes of the star. 
Because of their dominant first overtone content, these variables 
were hidden for a long period of time among the monoperiodic first 
overtone stars. In 1977 AQ~Leo was discovered as the first double-mode 
RR~Lyrae star by Jerzykiewicz \& Wenzel (1977). The first variable of 
this type in a globular cluster was discovered by Goranskij (1981). 
However, it was only in 1983 when the first systematic studies started 
in globular clusters (Cox, Hodson \& Clancy 1983), although there were 
suggestions that some of the first overtone variables in the appropriate 
period range with 'excessive scatter' might be actually double-mode stars 
(Sandage, Katem \& Sandage 1981). 

Less than a year ago Olech et al. (1999a, b, hereafter O99a,b) found 
that some first overtone stars in the globular clusters M5 and M55 
exhibit two frequencies, very closely spaced. They argued that the high 
period ratio strongly indicated the presence of nonradial modes in those 
stars. In the course of another recent study of the pulsation behavior of 
the RR~Lyrae stars in the Galactic bulge sample of the {\sc ogle} 
project, Moskalik (2000, hereafter M00) also found several variables 
with closely spaced frequencies. Furthermore, in a selectively chosen 
sample of the {\sc macho} RR~Lyrae inventory, Kurtz et al. (2000) also 
picked a few amplitude- and phase-modulated (Blazhko-type) variables 
among the first overtone stars. 

In the present paper we carry out a {\it systematic} study of the 
frequency spectra of 1350 first overtone stars of the {\sc macho} 
project for the Large Magellanic Cloud. The results of the investigation 
of a shorter segment of the same data set have already been briefly 
summarized by Kov\'acs et al. (2000). This is the first large-scale 
survey of the finer details of the temporal behavior of the variable 
star data of a microlensing survey. (We note however, that in a recent 
paper, Udalski et al. 1999, without going into the details, mention a 
similar massive analysis in a search for double-mode Cepheids in the 
Small Magellanic Cloud.) Due to the large size of the sample, this study 
also yields valuable statistics in respect of the occurrence rate of the 
various modal behaviors among first overtone stars. This, together with 
other information on the fundamental mode RR~Lyrae stars, supplies 
crucial observational data for understanding RR~Lyrae pulsation, and, 
in particular, the Blazhko phenomenon. 

Because of the various pulsation behaviors discovered among RR~Lyrae 
stars, we found it necessary to introduce new notation for the already 
known classical types. Throughout this paper we use RR0 for fundamental, 
RR1 for first overtone and RR01 for double-mode (fundamental \& first 
overtone), instead of the traditional notation of RRab, RRc and RRd. 
We will see that the new notation can be more successfully adapted to 
label new types and subtypes of variables.      

%
%
\section{THE DATA AND THE METHOD OF ANALYSIS}

The analysis presented in this paper utilizes the data obtained during 
the {\it first 6.5 years} of the {\sc macho} project. The primary 
selection of the sample was made through a cut in the color-magnitude 
diagram from the fields observed by the project (see 
http://wwwmacho.mcmaster.ca). The magnitude and color 
ranges were confined in the $18.5 \geq V \geq 20.0$ and 
$0.5 \geq V-R \geq 0.1$ intervals. This set was tested for 
variability by employing various statistics of the 
magnitude distributions of the individual objects. Next, the time 
series of the candidate variables were passed through a period search 
algorithm. In the final step of the primary selection the folded 
light curves were examined to filter out eclipsing binaries and 
other variables different from RR1 stars. Since the present sample 
contains variables only from 17 out of the 30 fields observed by 
the project, the sample is not complete. From the number of all 
variables (of any type) in the 30 fields and from the occurrence 
rate of RR1 variables in the 17 fields analyzed, we estimate the 
completeness to be 52\%. Details of the {\sc macho} image and 
photometry data are provided in Alcock et al.~(1999).  

Since our interest is focused only on the temporal behavior of the 
variables, in our primary survey we employ only the {\it instrumental 
`r' magnitudes}. However, when this color leads to dubious results, 
we also examine the frequency spectra corresponding to color `b'. 
Although the present RR1 sample contains about 1350 variables, some 
overlaps exist among the various fields both within this set and also 
with some of the 73 RR01 stars discovered by Alcock et al.~(1997, 
hereafter A97). 
%
%
%
\vskip 0mm
\centerline{\psfig{figure=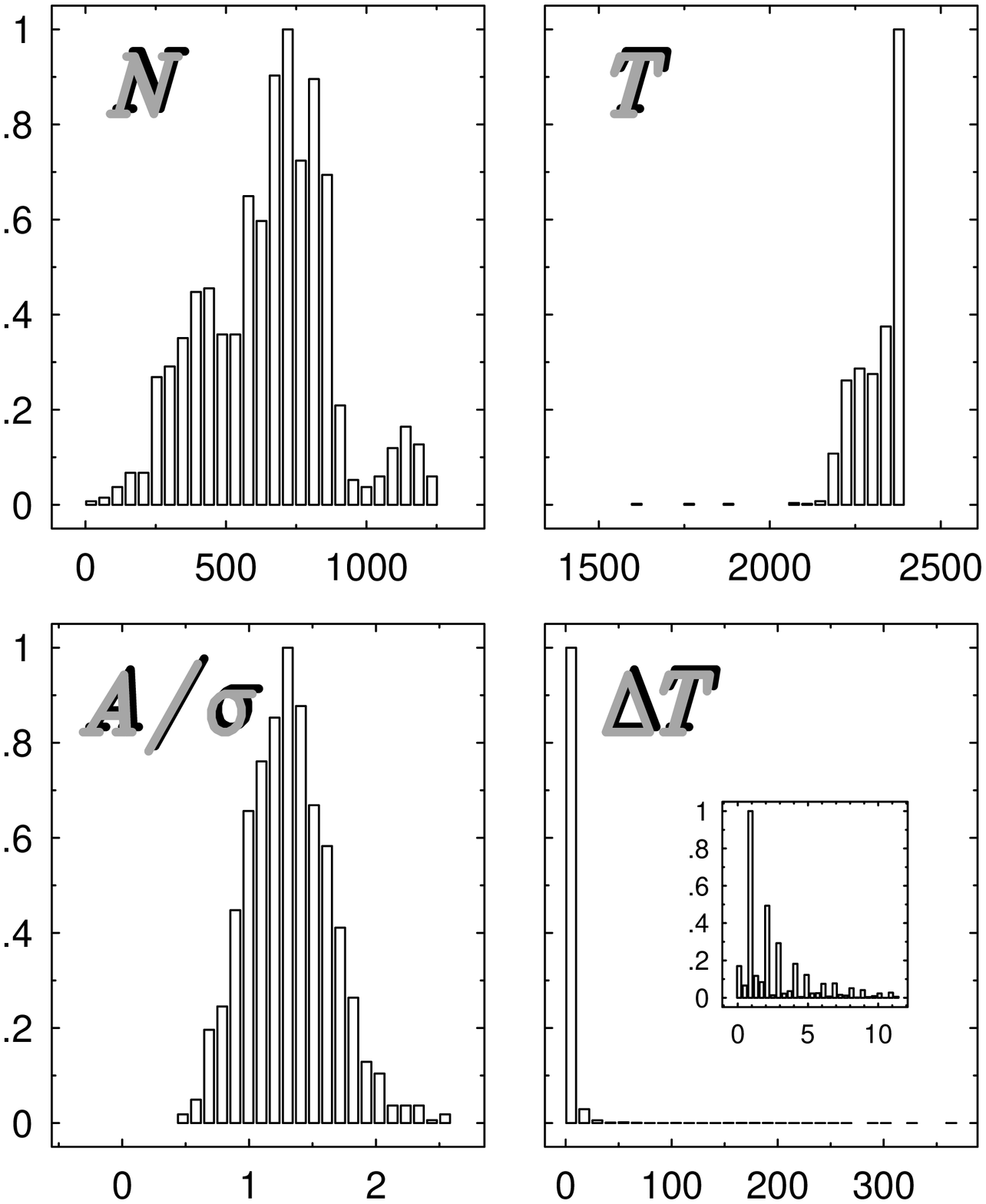,height=90mm,width=85mm}}
\vskip -5mm
\noindent 
{\footnotesize {\bf Fig. 1} -- Distribution functions of the various 
time-series parameters of the RR1 sample. {\it Upper left:} number of 
data points; {\it upper right:} total time span (in [days]); 
{\it lower left:} signal to noise ratio (see text); {\it lower right:} 
sampling time (in [days]). The inset shows the fine details of the 
$\Delta T$ distribution under 10 days. The highest peak of each 
distribution function is normalized to 1.} 
\vskip 3mm
\noindent
Further details on the data lengths, temporal distributions and signal 
to noise (S/N) ratios are given in Figure~1. We see that the most probable  
number of data points per variable is between 600 and 800 and is rarely less 
than 250, whereas for a considerable number of stars it exceeds 1000. The 
length of the total time span $T$ peaks at 2370~d which corresponds 
to the above mentioned 6.5~yr duration. Because of observational 
schedule, some of the variables cover a shorter time span, but very 
few of them extend for less than 6~yr. In the lower right corner we 
plot the distribution of the sampling time $\Delta T = t_{i+1} - t_i$. 
Because the LMC is circumpolar at Mount Stromlo, the data window is free 
from any long-term periodicity. The most significant problem comes from 
the daily sampling, but even this alias is tamed by the slight irregularity 
in the sampling rate as is exhibited by the small peaks between zero 
and 2~d in the distribution function. All these result in spectral 
windows which can be, in general, easily tackled as we will see in 
the subsequent sections. 

To characterize the noise properties of the light curves we define the 
signal to noise ratio $A/\sigma$ as the ratio of $A$, the amplitude of the 
first component of a 3rd order single-period Fourier fit, to $\sigma$, the 
standard deviation of the residuals of this fit. In the 
distribution function of $A/\sigma$, in the lower left corner of 
Figure~1, we see that the size of the noise is comparable to the main 
signal amplitude. Since only some 30\% of the total RR1 sample is 
multiperiodic (see later), the observed low S/N ratio is indeed due 
to the high noise, and not to the excess scatter attributed to 
secondary signal components. In some cases the light curves have 
substantial higher harmonic components, and, therefore, the total 
amplitudes might be twice of that of the first Fourier component, 
which results in a better S/N ratio. 
%
%
%
\vskip 5mm
\centerline{\psfig{figure=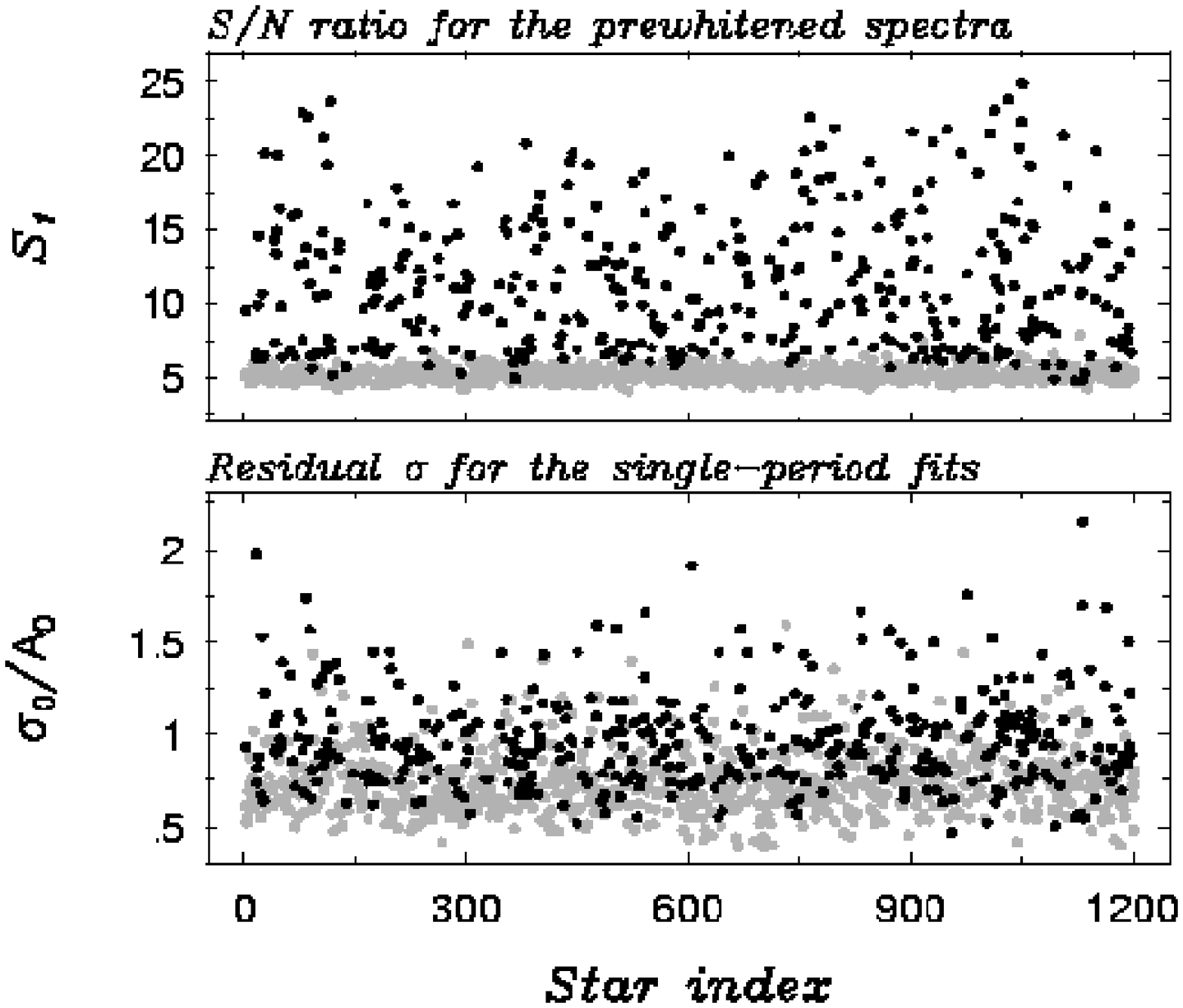,height=90mm,width=90mm}}
\vskip -6mm
\noindent 
{\footnotesize {\bf Fig. 2} -- 
Efficiency of searching for secondary signal component 
through Fourier frequency analysis of the prewhitened data 
({\it upper panel}) and through the residual scatter of the 
single-period fits ({\it lower panel}). {\it Gray dots} show the 
monoperiodic, {\it black dots} show the multiperiodic variables. Stars 
are listed by some registration index. 
} 
\vskip 3mm
\noindent
At such a low S/N ratio, when searching for secondary signal 
components, it is crucial to perform a frequency analysis combined with 
a {\it prewhitening} technique. During the prewhitening cycles we 
successively subtract the corresponding highest amplitude signal 
component and its first two harmonics from the time series. The fact 
that examination of the residual scatter in a single-period fit is 
{\it not} a decisive test for detecting secondary signal components  
in the low S/N case, is demonstrated in Figure~2. By using the observed 
data, in the lower panel we plot the relative residual scatter for the 
signal before the first prewhitening. Here $\sigma_0/A_0$ is the 
reciprocal of the same quantity plotted in Figure~1 (we use the zero 
subscript only to emphasize that the quantities refer to the data 
before the prewhitening cycles). We see that the multiperiodic stars 
are intermingled with the monoperiodic ones and only a few of them 
stand out relatively well from the rest of the sample. 

For characterizing the S/N ratio of the prewhitened spectra, the following 
quantity is introduced
%
\begin{eqnarray}
S_1 & = & {A_p - \langle A(\nu) \rangle\over \sigma_{A(\nu)}} \hskip 2mm . 
\end{eqnarray}
Here $A_p$ is the peak value in the amplitude spectrum, 
$\langle A(\nu) \rangle$ is the frequency-averaged value of the
spectrum, $\sigma_{A(\nu)}$ is the standard deviation of the amplitude 
spectrum in the given band width. The plotted values are obtained by 
considering the [0.5,5.5]~d$^{-1}$ frequency band. We see that most 
of the detections are rather safe, with confidence limits better than 
$6\sigma$. It is clear that in future automatic search for 
multiperiodic variables statistics $S_1$ could be very helpful.  

The frequency analysis is carried out by employing a standard Discrete 
Fourier Transformation (DFT) for unequally spaced data (see e.g., 
Deeming 1975). The input time-series of the DFT is obtained by 
subtracting the average from the original data set and omitting 4\% 
of the data points which have the largest deviations from the average. 
We take the [0.5,5.5]~d$^{-1}$ frequency band for the primary survey  
and the [0.0,6.0]~d$^{-1}$ band for the detailed analysis of the multiperiodic 
cases. All spectra in the above bands are calculated with 60000 and 80000 
frequency steps, respectively. To avoid plotting unimportant features and 
unnecessarily large arrays, we employ {\it economic plotting}, which 
means that the plotted values correspond to the maximum amplitudes in 
the bins of the chosen bin resolution of the total frequency band. 
The number of bins in our case is 2000. In each prewhitening cycle 
the location of the highest peak in the given band width is used to 
calculate a more accurate estimation of its frequency. This is done 
by a least-squares technique, by searching for the minimum dispersion 
around a 3rd order single-period Fourier-sum. Up to three prewhitening 
cycles are performed in the multiperiodic cases. All spectra  
and folded light curves are {\it visually} inspected and then 
{\it classified according to the prewhitened spectra}. As we see from 
the statistics of $S_1$ in Figure~2, in principle one could pick up most 
of the multiperiodic variables on the basis of the $\approx 6\sigma$ 
criterion. However, at this stage of the analysis of a still `reasonably' 
low number of variables, the `manual' search is preferred in order to 
ensure that no interesting cases are missed because of slightly 
marginal detectability. Finally, in the class of variables showing  
period change, we employ a very simple time-dependent Fourier analysis, 
which consists of a series of single-period Fourier fits to the adjacent 
parts of the time series.  

%
%
\section{RR01-TYPE STARS: VARIABLES WITH THE FUNDAMENTAL \& FIRST 
OVERTONE MODES}

This class of variables is the simplest to detect in the present 
\hfill data \hfill set. 
The \hfill standard \hfill prewhitening \hfill method
%
%
%
\vskip 0mm
\centerline{\psfig{figure=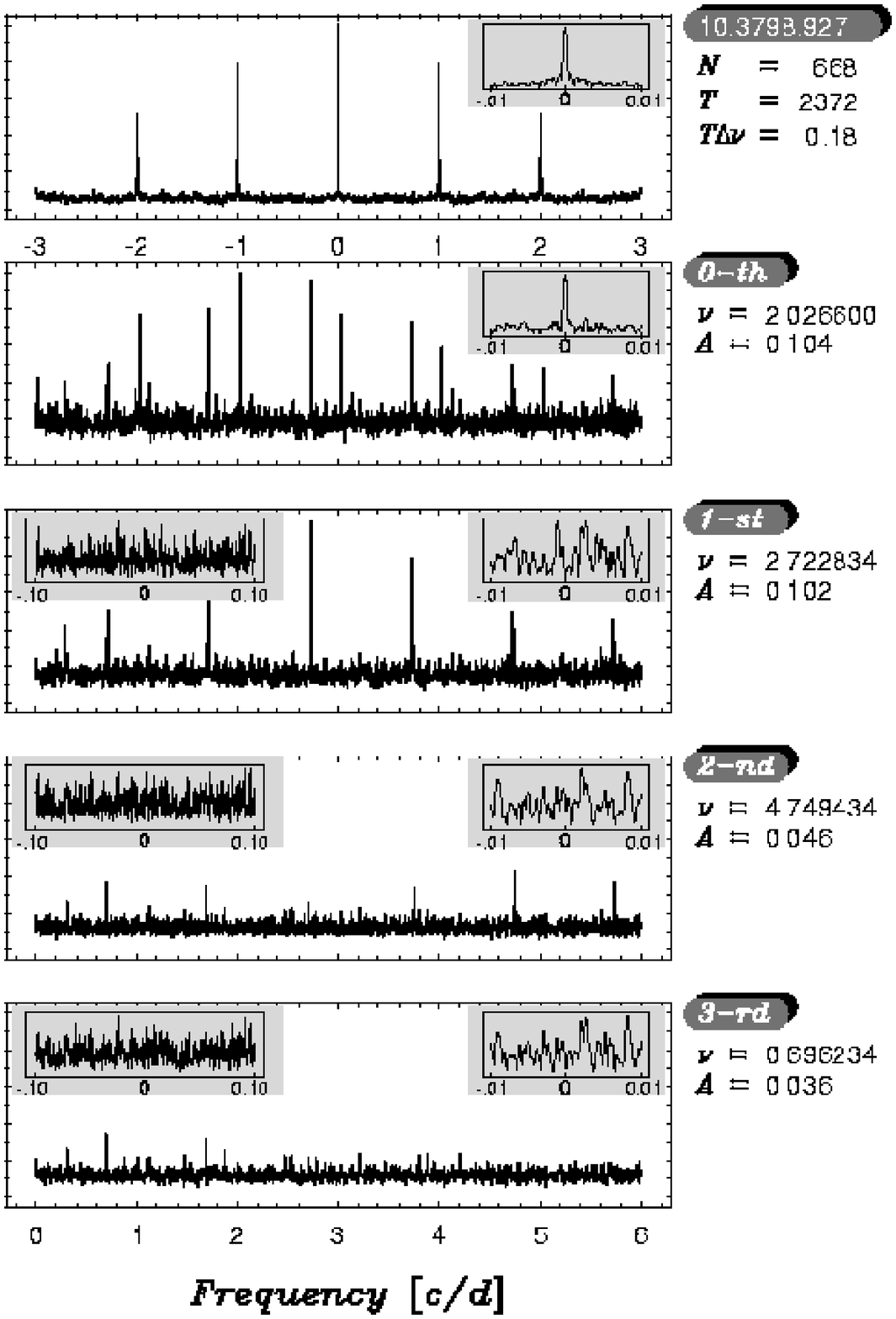,height=130mm,width=90mm}}
\vskip 0mm
\noindent 
{\footnotesize {\bf Fig. 3} -- 
Example of the detection of a double-mode (RR01) variable. 
The upper panel shows the spectral window. Inset displays the fine 
structure around the main peak. Label shows the {\sc macho} identifier, 
number of data points ($N$), length of the total time span 
($T$ in [days]) and frequency resolution ($T\Delta\nu$) of this and 
all the following wide frequency band spectra. Subsequent panels 
show the amplitude spectrum of the original data and the results of 
the corresponding prewhitening cycles (see Sect. 2 for details). 
Insets show the fine structure of the spectra centered around the 
main peak of the spectrum of the original data. Labels show the 
prewhitening order, the frequency and the amplitude (in [d$^{-1}$] 
and in {\sc macho} `r' magnitude, respectively) of the main peak of 
the corresponding prewhitened spectra. The wide frequency band spectra 
are normalized relative to the highest peak in the spectrum of the 
original data. The spectra shown in the insets are normalized separately 
to the highest peaks in the narrow frequency band of the corresponding 
spectra. As it is described in the text, economic plotting is employed 
in all cases.
} 
\vskip 5mm
\noindent
based on the maximum peak subtraction 
almost always works, although we use a wide frequency band and do not 
restrict the search to some `proper' period ratio regime. 
This latter condition is applied only in one or two cases from the 
156 successful searches. (We note, however, that 48 stars from this set 
are either 
doubly identified or have already been discovered by A97, using a reduced 
and different data set.) In many cases the lowest order combination 
frequencies $f_1 - f_0$ and $f_1+f_0$ are also identified (the harmonics 
are automatically subtracted by our 3rd order monoperiodic fits). Figure~3 
illustrates the case when these combination frequencies are easily 
visible. We draw attention to the following important features: 
(a) the spectral window contains only the integer d$^{-1}$ aliases, and 
is free from any long-term alias problems (this property is true for 
almost all time series); the 1~d$^{-1}$ aliases are sufficiently small 
and in most cases do not hamper the identification of the maximum peak 
in the spectra; (b) the combination frequencies are very well reproduced 
from $f_0$ and $f_1$, which is a strong indication of the physical nature 
of the frequencies found.    
The basic parameters of the previously unknown RR01 variables are 
given in Table 1\footnote{Tables 1 -- 6 are presented after the 
{\sc references}}. (The amplitudes $A_1$ and $A_0$ refer to the first 
overtone and fundamental modes, respectively, and are obtained 
through monoperiodic Fourier fits during the prewhitening procedure. 
The coordinates refer to the year 2000.) Together with the already 
published variables by A97, there are 181 RR01 stars known 
in the LMC. This number of variables is about three times of the total 
number of RR01 stars presently known in globular clusters, galaxies 
and in the Galactic field. 
%
%
%
\vskip 0mm
\centerline{\psfig{figure=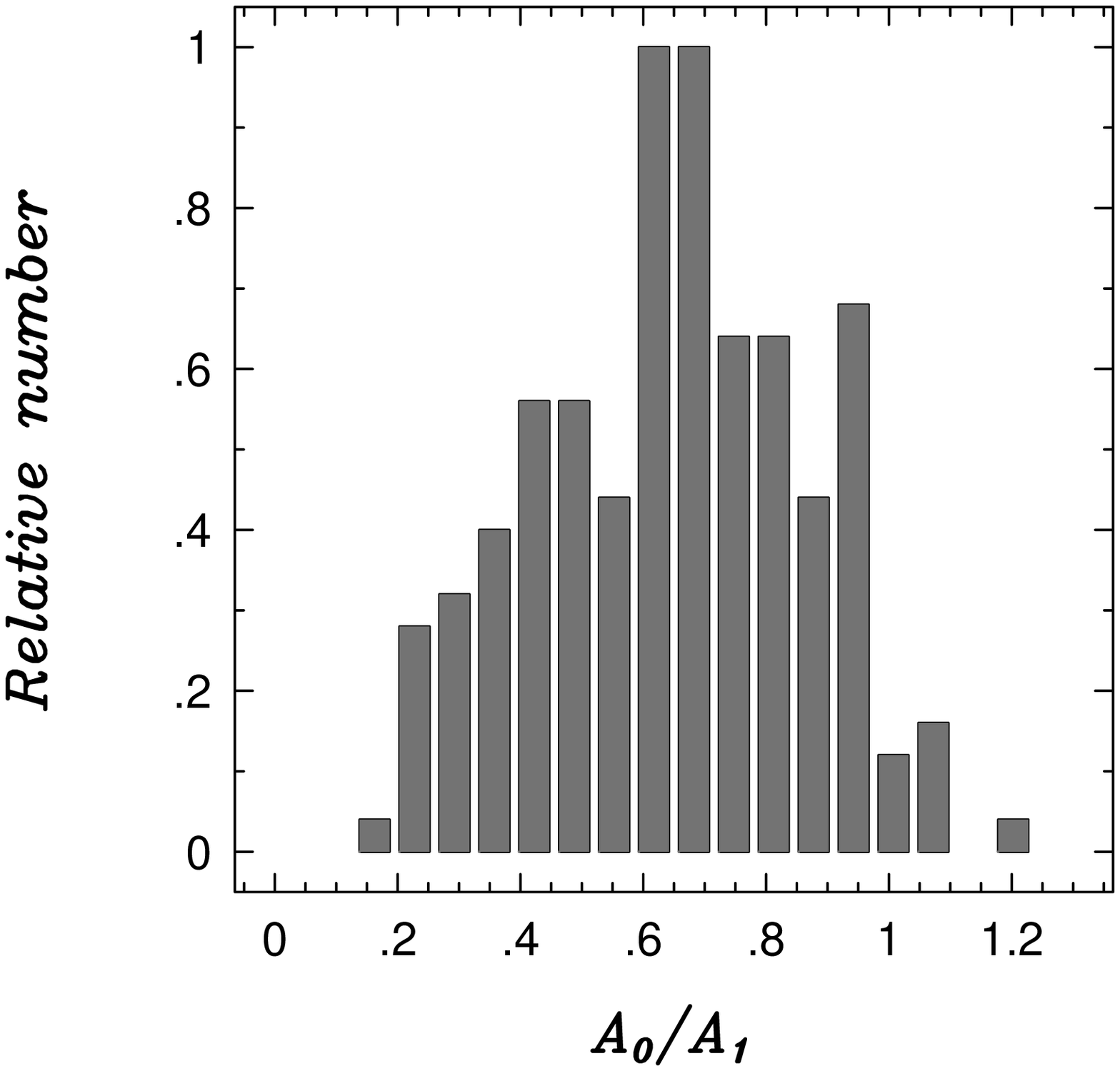,height=80mm,width=90mm}}
\vskip -0mm
\noindent 
{\footnotesize {\bf Fig. 4} -- 
The distribution of the fundamental to first overtone 
amplitude ratios for all the 181 RR01 stars in the LMC.  
} 
\vskip 3mm
\noindent
When plotted either on the period--amplitude or on the period--apparent 
magnitude diagrams, LMC RR01 stars do not seem to be distinguishable from 
the single-mode RR1 stars. They occupy a relatively narrow period range 
of $0.33$~d~$< P_1 <$~$0.41$~d (there is only one star with $P_1 = 0.43$~d). 

As it is shown in Figure~4, the ratio of the amplitudes of the two 
modes shows a wide range, but the fundamental mode amplitude $A_0$ 
rarely exceeds that of the first overtone. Apparently there is a lower 
cut in the $A_0/A_1$ ratio at 0.2, because we do not observe lower 
ratio than this, although our detection limit is below this value 
(see the tests for the Blazhko variables in Sect. 5). 
%
%
\vskip  2mm
\centerline{\psfig{figure=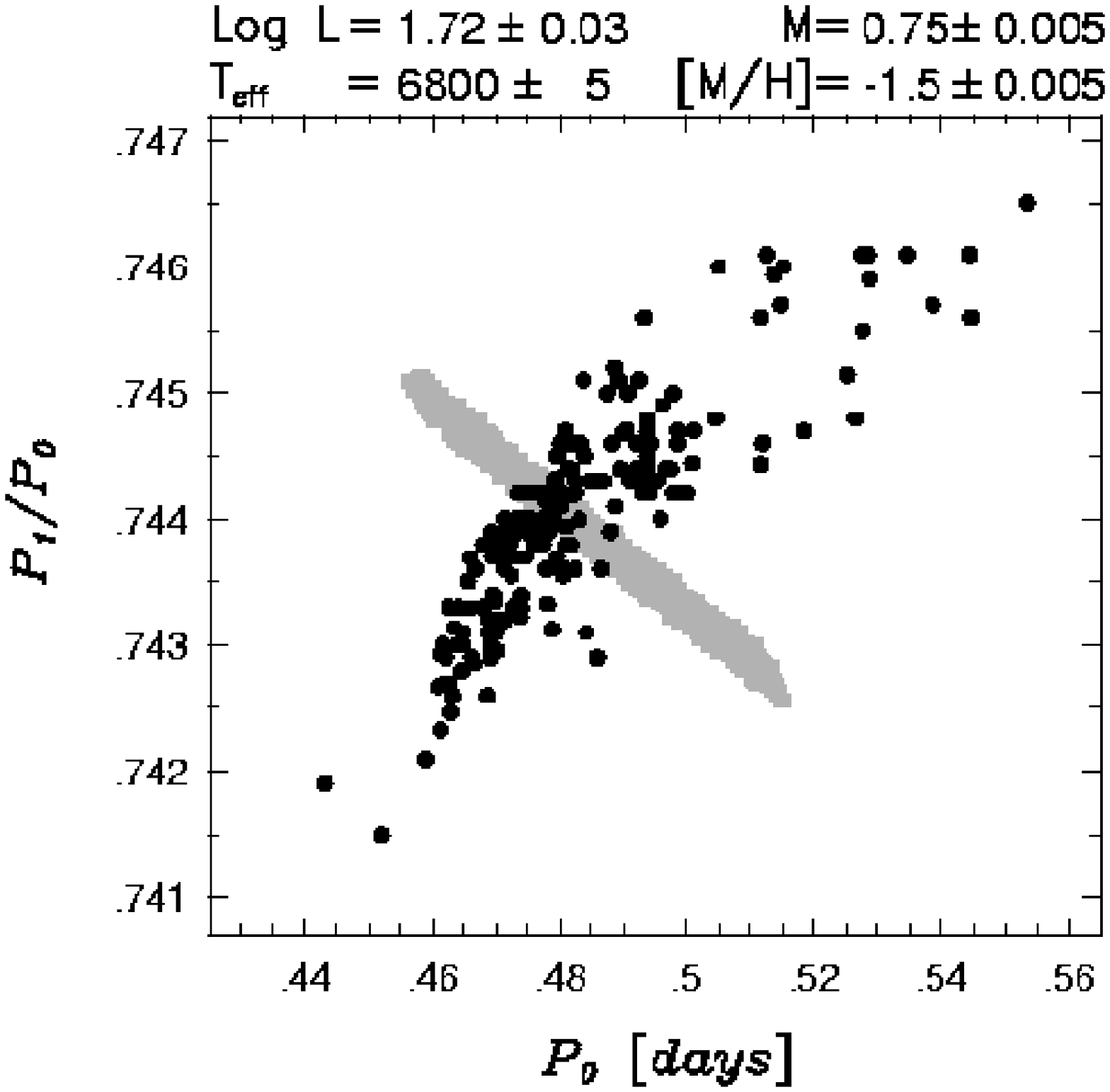,height=90mm,width=90mm}}
\vskip -3mm
\noindent 
{\footnotesize {\bf Fig. 5} -- 
Position of the LMC RR01 stars in the $P_0\rightarrow P_1/P_0$ 
diagram. For comparison, we also show by gray-colored background the 
area occupied by models selected in the parameter regime shown in the 
header. The models of Kov\'acs \& Walker (1999) are used with the 
{\sc opal'96} opacities. The Hydrogen abundance $X$ is fixed at $0.76$ 
and the adopted value of the solar metallicity $Z_{\odot}$ is $0.02$. 
Mass and luminosity are given in solar units. 
} 
\vskip 3mm
\noindent
It is well known that the $P_0\rightarrow P_1/P_0$ diagram (Petersen 1973) 
carries important information about the metal content and mass of the RR01  
stars (Kov\'acs, Buchler \& Marom 1991). In Figure~5 we plot this diagram 
for all known LMC RR01 stars, together with the region of the models 
occupied at some fixed metallicity and mass. It is seen that the constant 
mass and metallicity assumption is not applicable to the whole RR01  
population of the LMC. Assuming that either the mass or the metal content 
is the sole agent for the observed topology, we get rough estimates of 
$\pm 0.15 M_{\odot}$ and $\pm 0.3$~dex, respectively, necessary to cover 
the total ranges of periods and period ratios. There is also a dependence 
on $L$ and $T_{eff}$ which influences the applicable average values of 
mass and [M/H] (in the case displayed we use the so-called `brighter' 
luminosity scale -- see Kov\'acs \& Walker 1999). It is obvious that 
the proper treatment of this problem requires additional information, 
most importantly accurate abundance values of the LMC RR01 variables. 
For further discussion of this problem we refer to A97, Popielski 
\& Dziembowski (2000) and Clementini et al. (2000).

%
%
\section{RR1$-\nu1$-TYPE STARS: VARIABLES WITH 2 CLOSELY SPACED FREQUENCIES}

These variables represent a special type of double-frequency 
pulsation. The frequency ratios are always larger than 0.95 and could  
be as large as 0.999. In Figure~6 we show an example for the patterns 
obtained in the case of medium frequency separation. It is clear 
that in the frequency band analyzed, there are no other components 
except for the 2 closely spaced ones we identified. (As it is seen 
in the third panel, the nearly equidistant triplet structure of 
the low-frequency aliases is due to the power leakage from negative 
frequencies.) \hfill Other \hfill solutions, \hfill such 
%
%
\vskip -0mm
\centerline{\psfig{figure=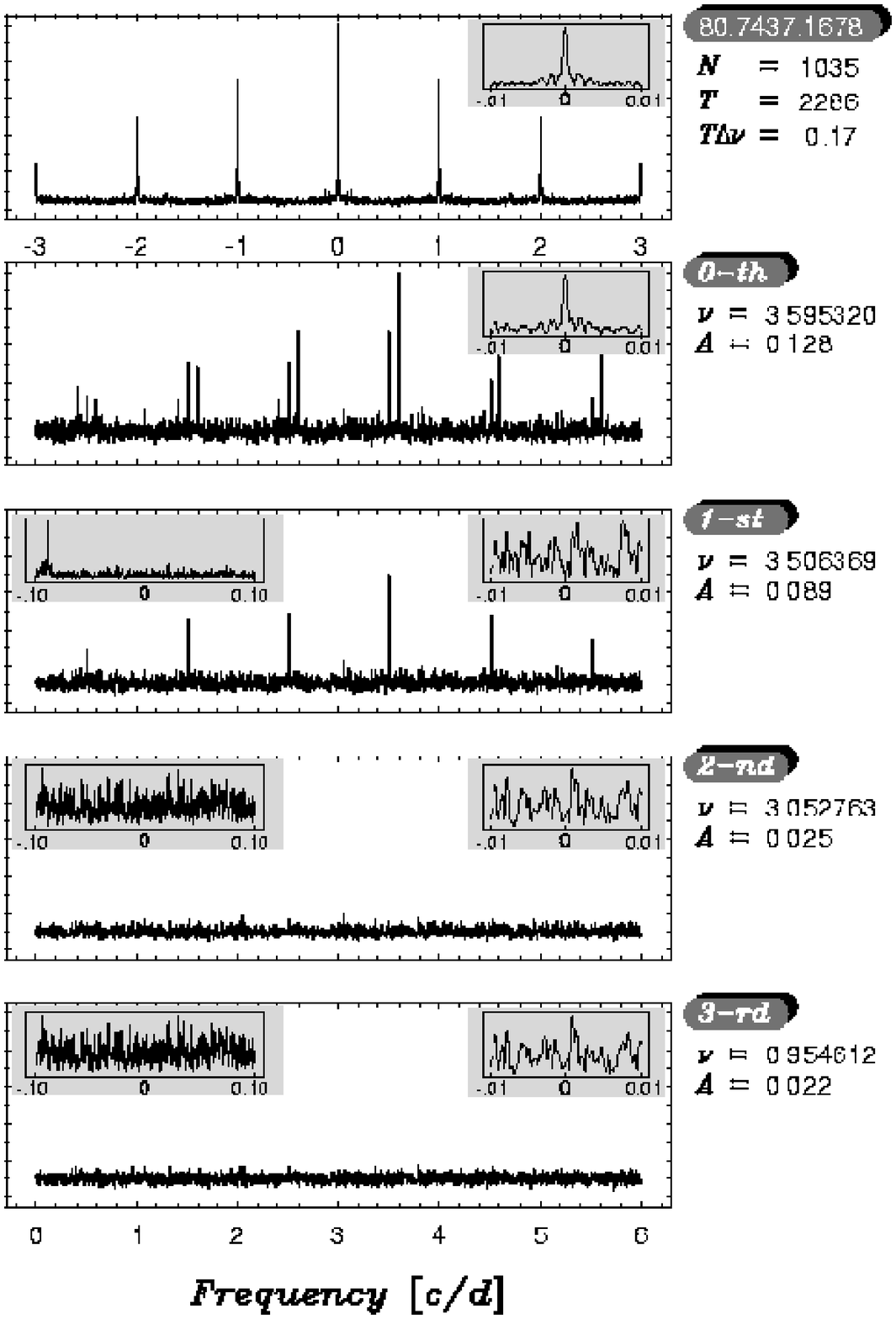,height=130mm,width=90mm}}
\vskip -0mm
\noindent 
{\footnotesize {\bf Fig. 6} -- 
Detection of an RR1$-\nu1$ variable (RR1 star with 2 closely 
spaced frequencies). Notation is the same as in Figure~3.
} 
\vskip 3mm
\noindent
as the RR01 frequency pattern are not 
possible, both because of the fairly clear alias patterns obtained 
during the calculation of the present solution, and also because 
on physical grounds (i.e., we would have obtained `bad' period ratios 
with the RR01 assumption). Unlike in the case of Galactic bulge RR0$-\nu1$ 
variables analyzed by M00, we do not find traces of the combination 
frequencies $\nu_1-\nu_0$ and $\nu_1+\nu_0$. There is only one variable 
with frequency separation smaller than 0.01~$d^{-1}$ (see Figure~7). 
The status of this star is slightly ambiguous, because the `b' data 
suggest the presence of a symmetric frequency pattern (see next section), 
although the feature is sufficiently hidden in the noise, similarly to 
the case of the `r' data shown in the figure.

The main properties (frequency $\nu_0$ and amplitude $A_0$ of 
the main -- i.e., higher amplitude -- component, frequency separation 
$\nu_1-\nu_0$, and the ratio $A_1/A_0$ of the secondary and 
main amplitudes) of the 24 RR1$-\nu1$-type variables are listed in Table 2. 
Most of these stars have frequency separations between 0.01 and 
0.1~d$^{-1}$ and only a few of them exceed these limits. 63\% of the 
variables have negative frequency differences (i.e., the secondary, 
lower amplitude component has smaller frequency than the main component). 
In the case of positive frequency differences the secondary amplitudes 
are almost always quite comparable with those of the main components. 
It is also seen that the period regime in which RR1$-\nu1$-type pulsation 
occurs, is fairly wide. 
%
%
\vskip -0mm
\centerline{\psfig{figure=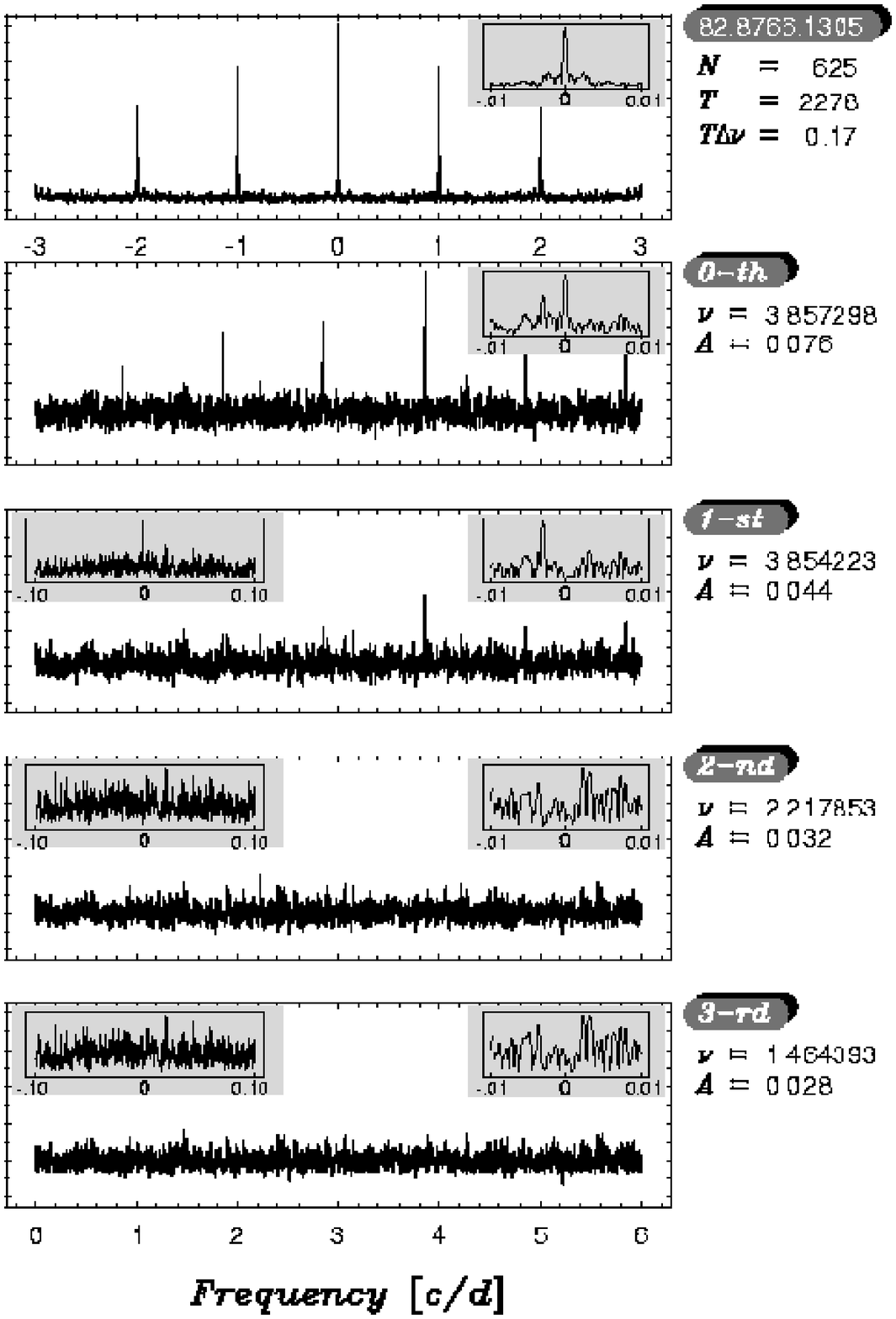,height=130mm,width=90mm}}
\vskip -0mm
\noindent 
{\footnotesize {\bf Fig. 7} -- 
RR1$-\nu1$ variable with small frequency separation. For notation, 
see Figure~3.
} 
\vskip 3mm
\noindent
For a closer inspection of the overall properties of the RR1$-\nu1$ 
spectra, in Figures 8 and 9 we show a representative sample of this 
type of variables. In many cases the secondary components are 
very easily visible even without prewhitening. In some prewhitened 
spectra of sufficiently low noise level, one may suspect the presence 
of additional low-amplitude components. Specially interesting are 
those in which the suspected component forms a symmetric triplet with 
the main and secondary components. As an example, we recall the case 
of 82.8766.1305 (see note to Table 2), where the `b' data support the 
existence of such a symmetric frequency pattern. On the other hand, 
a closer examination of the suspected third component for 2.5266.3864 
reveals that it is asymmetric and becomes very small after the 
second prewhitening, whereas for 80.6352.1495 the very marginal 
possible third component indicated in Figure~9 completely disappears. 
These conclusions are also supported by the analysis of the `b' data. 
We return to the problem of hidden symmetric frequency patterns in the 
next section during the discussion of the Blazhko variables. 

%
%
\vskip -0mm
\centerline{\psfig{figure=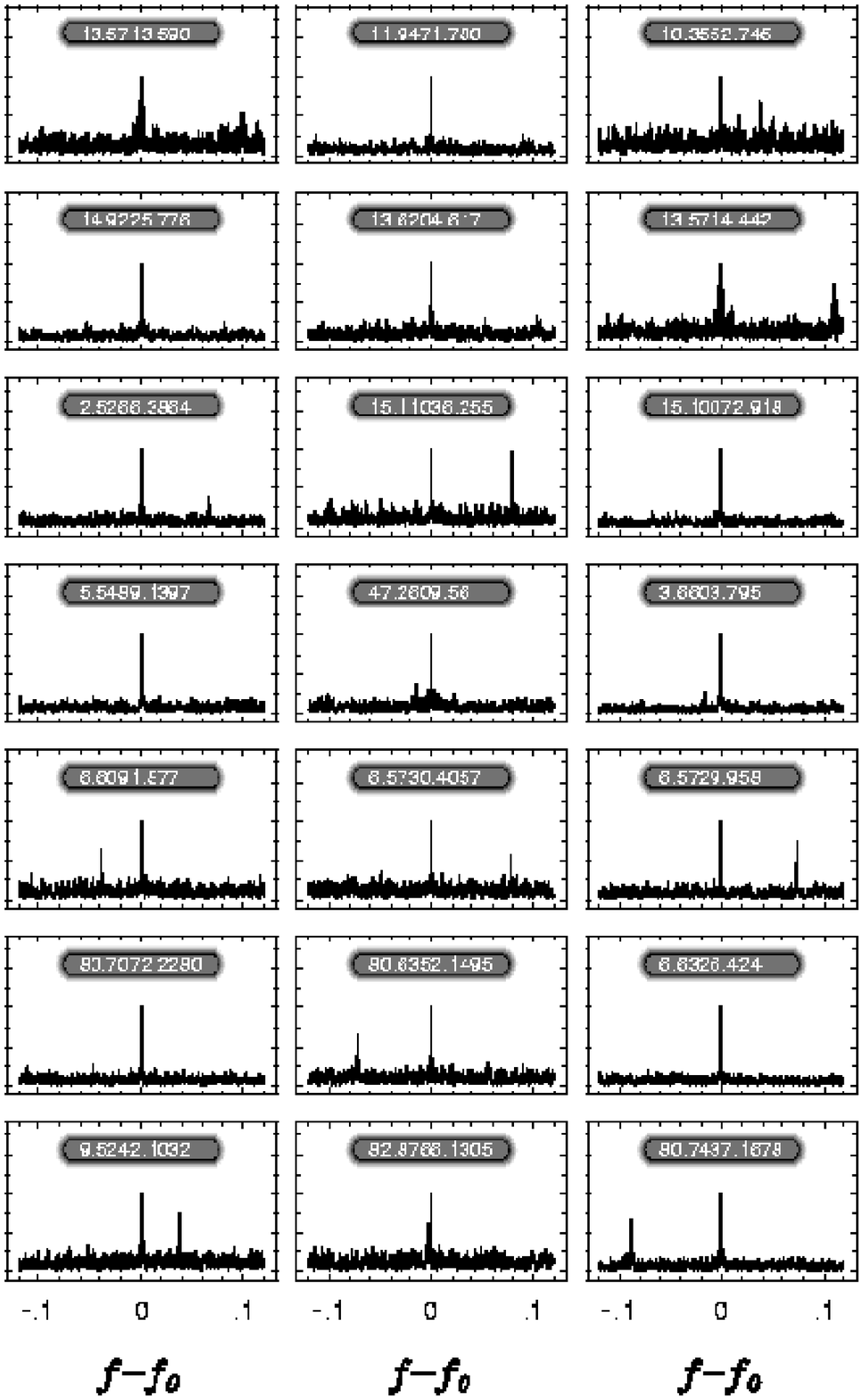,height=140mm,width=110mm}}
\vskip -0mm
\noindent 
{\footnotesize {\bf Fig. 8} -- 
Amplitude spectra of a sample of RR1$-\nu1$ stars without 
prewhitening. Each spectrum is normalized to the same level by 
the corresponding highest peak.
} 
\vskip 7mm
\noindent
As we have already mentioned in Sect. 1, parallel with the present 
study, RR1$-\nu1$-type variables have also been discovered in other 
stellar systems. There are 3 RR1$-\nu1$-type stars in the globular 
cluster M55, 1 in M5 (O99a,b) and 2 in the {\sc ogle}-sample of 
66 RR1 stars of the Galactic bulge (M00). It is worthwhile to mention 
that for four of these variables the frequency difference is negative 
in agreement with most of the values in our sample. In addition to the 
two RR1$-\nu1$-type stars, M00 found similar type of stars in the 
RR0 population. From the 149 stars he identified 11 variables with 
two closely spaced frequencies. It is interesting to note that 9 of 
these stars have frequency spacings with opposite sign to that of 
the majority of the RR1$-\nu1$ stars. The authors of both of these 
papers argue strongly that these discoveries are the first observational 
proofs for the existence of {\it nonradial modes} in RR~Lyrae stars. 
Indeed, it is difficult to escape this conclusion, especially in the 
light of the recent theoretical works indicating the excitation of 
nonradial modes in RR~Lyrae stars (Van Hoolst, Dziembowski 
\& Kawaler 1998; Dziembowski \& Cassisi 1999). 

%
%
\vskip -0mm
\centerline{\psfig{figure=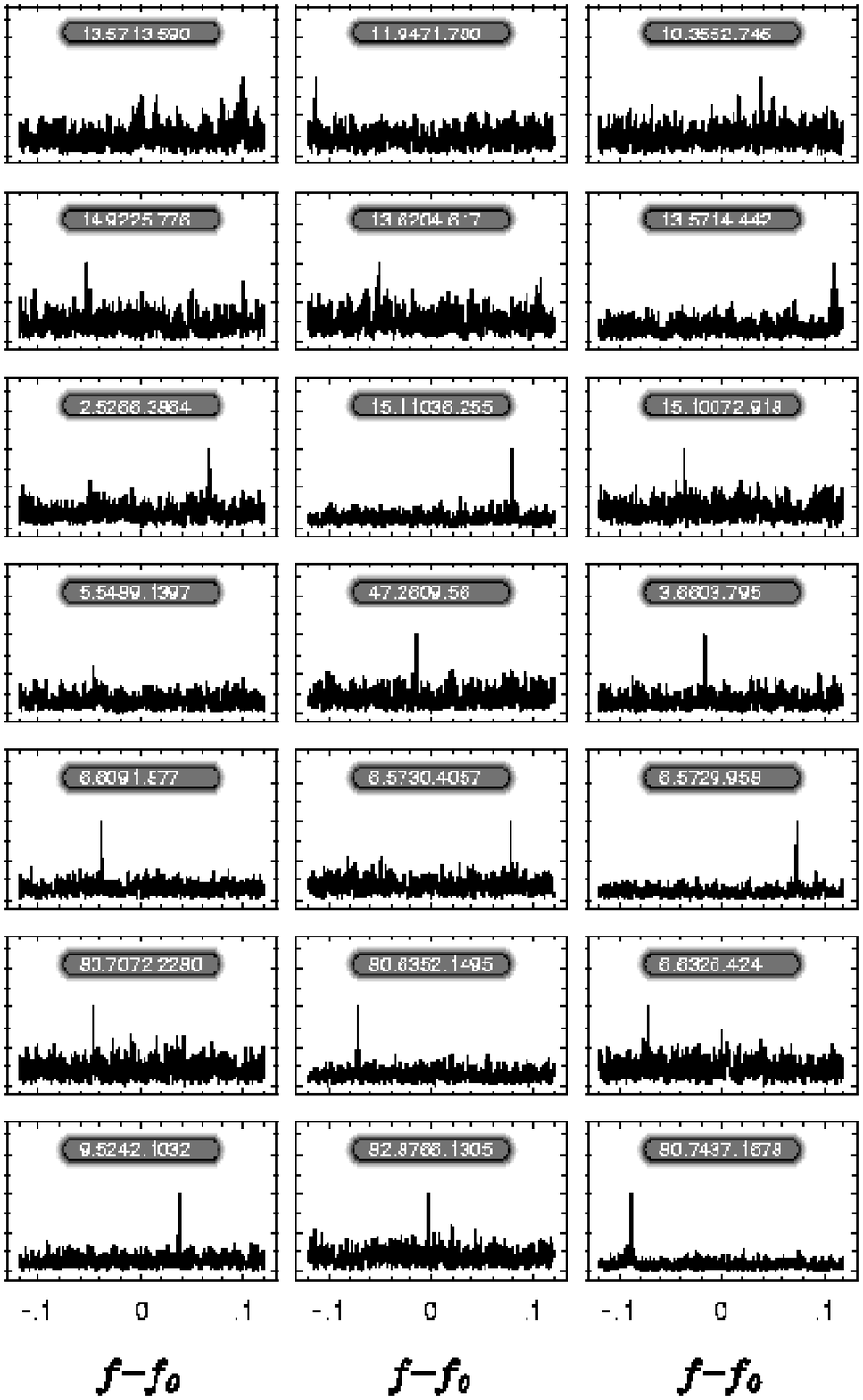,height=140mm,width=110mm}}
\vskip -0mm
\noindent 
{\footnotesize {\bf Fig. 9} -- 
Amplitude spectra of a sample of RR1$-\nu1$ stars for the 
data after the first prewhitening. Each spectrum is normalized to the same 
level by the corresponding highest peak. 
} 
\vskip 10mm
%
%
 
%
%
\section{RR1$-BL$-TYPE STARS: VARIABLES WITH 3 SYMMETRICALLY SPACED CLOSE 
FREQUENCY COMPONENTS}

Fourier analyses of Galactic field RR0 stars with Blazhko effect 
(Borkowski 1980; Smith et al. 1994, 1999; Kov\'acs 1995; Nagy 1998; 
Szeidl \& Koll\'ath 2000) revealed a very simple frequency pattern 
for these stars. The spectra constitute a sequence of 
{\it equidistant triplets}, centered around the fundamental mode 
frequency and its harmonics. This structure can be observed up to 
the 5th -- 7th harmonics. The frequency spacing corresponds to the 
modulation (Blazhko) period. The amplitudes of the modulation 
components around the harmonics are 10--30\% of the corresponding 
harmonics and are, in general, not symmetric. Their contributions  
add up in the directly observable amplitude and phase changes and 
result in considerable amplitude variations of 30--50\% (Szeidl 1988). 
  
Before the present analysis (see also Kurtz et al. 2000) it was 
unclear whether Blazhko-type modulation was also present in RR1  
stars. Interestingly, we find that this behavior does also occur 
in RR1 stars; however, its incidence is much lower than among 
RR0 stars. We discovered altogether 28 variables showing Blazhko-type 
frequency patterns. The details of the prewhitening procedure 
for such a variable are shown in Figure~10. Again, here we draw the 
attention to the flat spectrum after prewhitening by all three peaks. 
Equal frequency spacing is also well displayed. Although most of the 
variables have modulation frequencies similar to the example shown, 
7 variables have rather small ones, corresponding to modulation 
periods longer than 600~d . In Figure~11 we show an example for such 
a variable. It is seen that despite the very small separation, the 
components are still clearly resolved due to the long baseline of 
the data. 
%
%
\vskip -0mm
\centerline{\psfig{figure=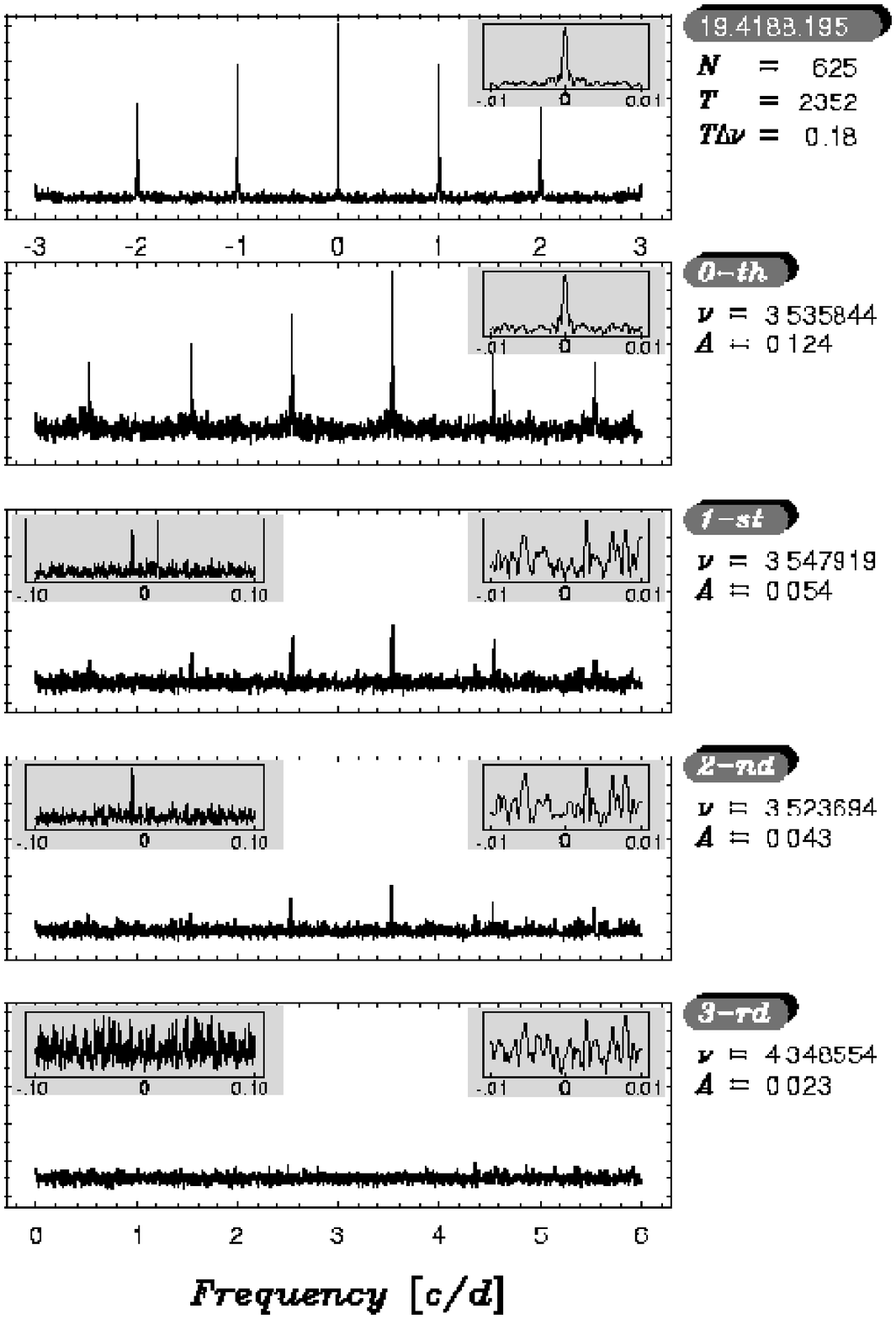,height=130mm,width=90mm}}
\vskip -0mm
\noindent 
{\footnotesize {\bf Fig. 10} -- 
Detection of a first overtone Blazhko variable 
(RR1 star with 3 symmetrically spaced close frequencies). Notation is the 
same as in Figure~3.
} 
\vskip 10mm
\noindent
The results of the analysis for the RR1$-BL$-type variables are 
summarized in Table 3. The quantities corresponding to the highest 
peak in the frequency spectrum before prewhitening are labeled with 
zero subscripts. The $+$ and $-$ signs denote components which 
have, respectively, larger and smaller frequencies than the main 
component (i.e., $\Delta f_{\pm}=\nu_{\pm}-\nu_0$). We see that except 
for 11.9355.1380, in all cases the main component is also the center 
of the triplet. The lowest and highest detected modulation levels are 
12\% and 82\%, respectively.
%
%
%
\vskip -0mm
\centerline{\psfig{figure=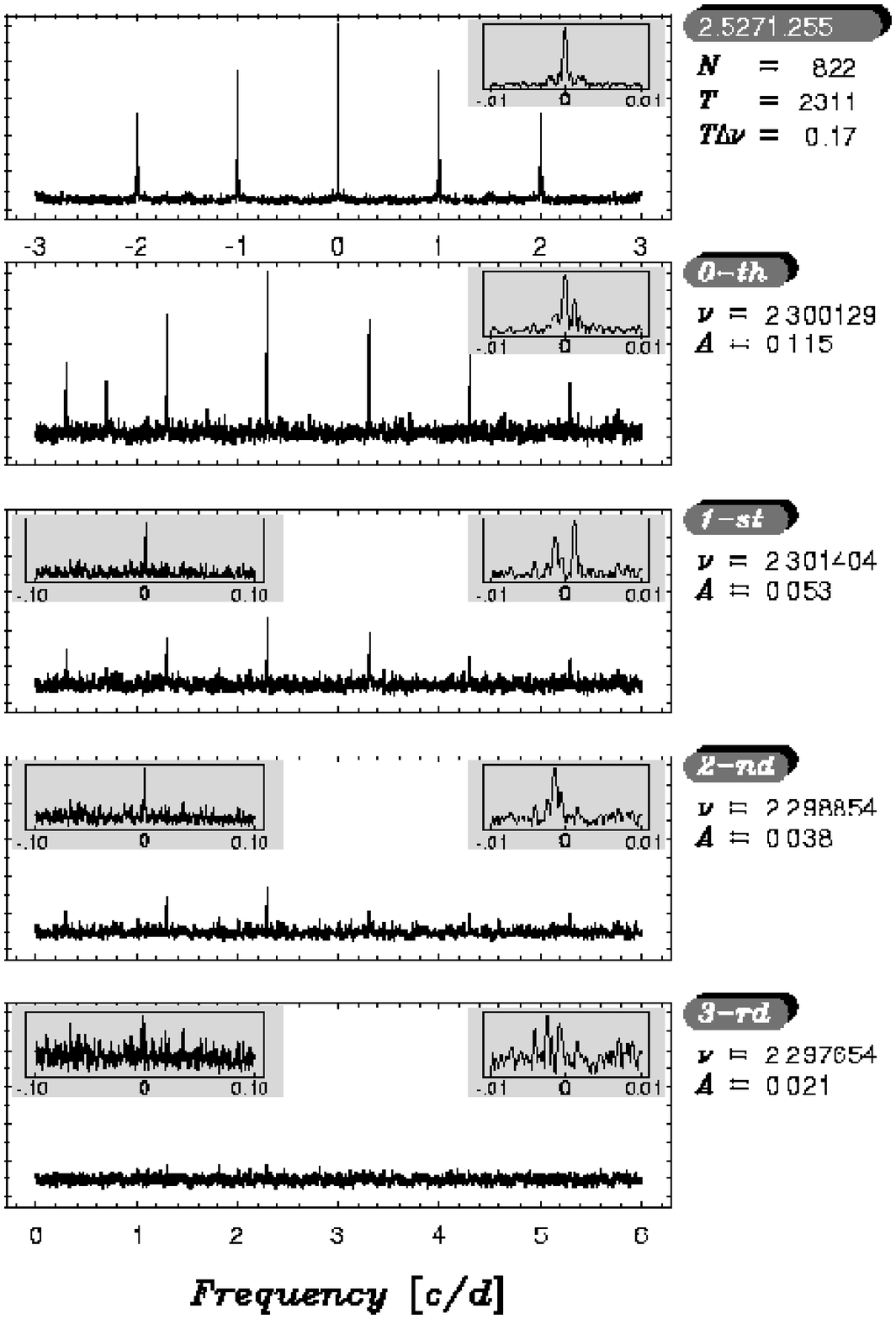,height=130mm,width=90mm}}
\vskip -0mm
\noindent 
{\footnotesize {\bf Fig. 11} -- 
Blazhko variable with very small frequency separation. For 
notation, see Figure~3.
} 
\vskip 3mm
\noindent
For an overview of the patterns of the RR1$-BL$-type frequency spectra, in 
Figures 12 and 13 we display the spectra of the data before and after the 
first prewhitening. In some cases we see remnants precisely at the 
$\nu_0$ component. These are either due to long-period modulations 
not resolved on the present plotting scale, or to period changes of 
the main components. Furthermore, in other cases, the presence of 
additional close component(s) can also be suspected (e.g., 82.8765.1250). 

In the following we address two questions which are important for the 
physical models of the Blazhko phenomenon: (a) What is the statistical 
significance of the slight deviations from equidistant frequency 
spacing? (b) How small is the lowest observable level of modulation 
in the present data set? 

Problem (a) is tested in the following way. For each variable to be 
tested for the significance of the equidistant frequency spacing, 
we generate a synthetic signal by using the average of the observed 
frequency distances and the corresponding Fourier decomposition 
obtained from fitting this equidistant frequency triplet to the data. 
Then Gaussian noise is added to this signal with the standard 
deviation of the residuals of the Fourier fit of the original data 
mentioned above. After analyzing many realizations of the so-obtained 
time series, we check the statistical properties of the calculated 
frequency distances. The deviation from 
%
%
%
\vskip -0mm
\centerline{\psfig{figure=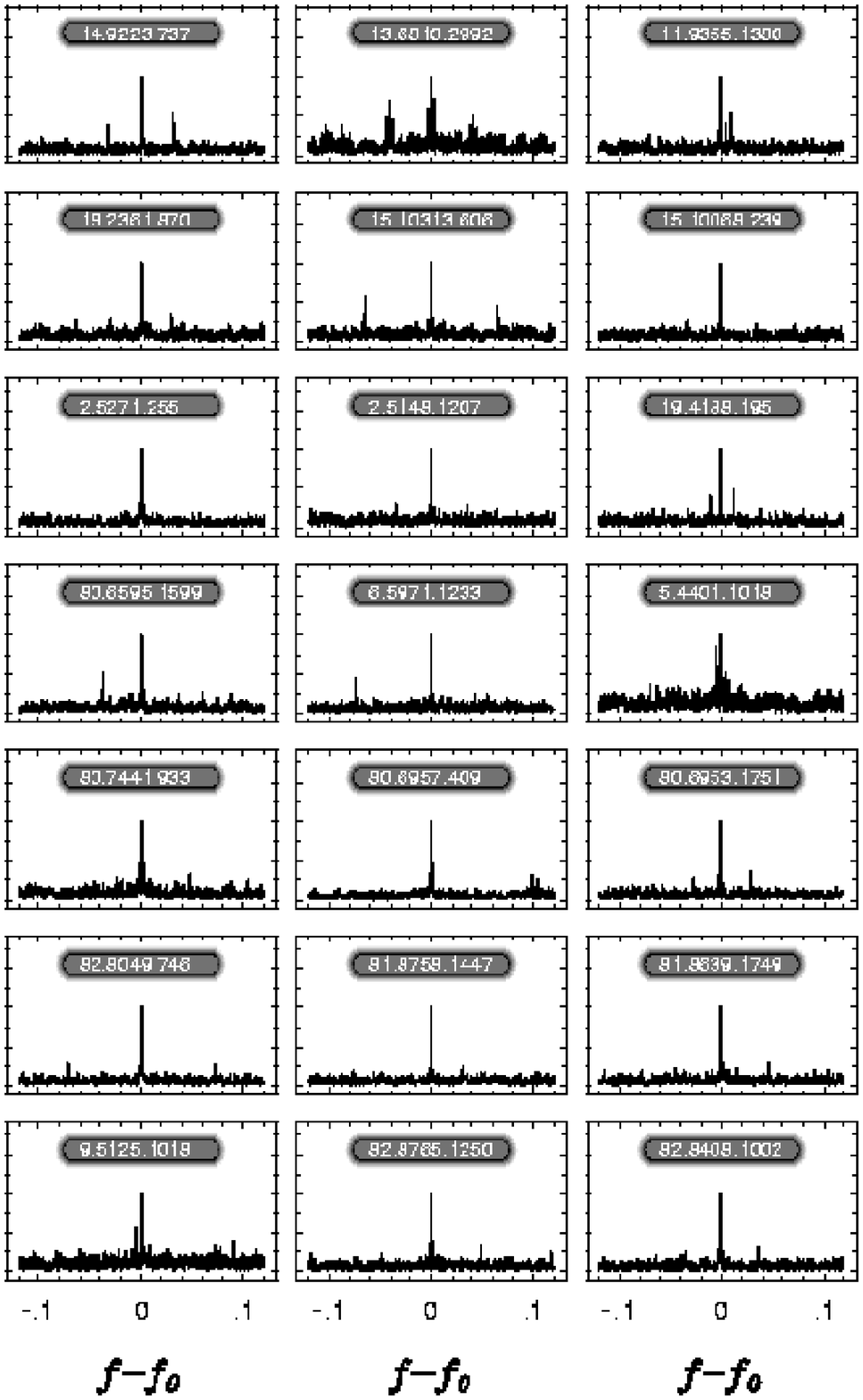,height=140mm,width=110mm}}
\vskip -0mm
\noindent 
{\footnotesize {\bf Fig. 12} -- 
Amplitude spectra of a sample of Blazhko stars without 
prewhitening. Each spectrum is normalized to the same level by 
the corresponding highest peak.
} 
\vskip 3mm
\noindent
the equidistant pattern is 
measured by the absolute value of the difference between the two 
spacings, i.e.,      
%
\begin{eqnarray}
\delta f = \vert \Delta f_{+} + \Delta f_{-} \vert  \hskip 2mm . 
\end{eqnarray}
In Figure~14 we show the empirical probability distribution of 
$\delta f$ for one of the variables in Table 3. We see that the 
observed deviation is well within the probability limit expected 
from the effect of observational noise. The same conclusion is drawn 
from the tests performed with other variables. 

Turning to problem (b), we see from Table 3 that the modulation level  
$A_{\pm}/A(\nu_0)$ for most of the RR1$-BL$ variables is larger than 
20\%. We have only one variable for which both components are under this 
limit. The question of lowest modulation level is important from the 
point of view of the relation between the Blazhko RR0 and RR1 stars, 
because the possibility of the existence of RR1 stars with low-level 
modulation would increase the number of Blazhko RR1 stars, thereby 
decreasing the gap between the occurrence rates of the two classes 
of Blazhko variables.  
%
%
%
\vskip -0mm
\centerline{\psfig{figure=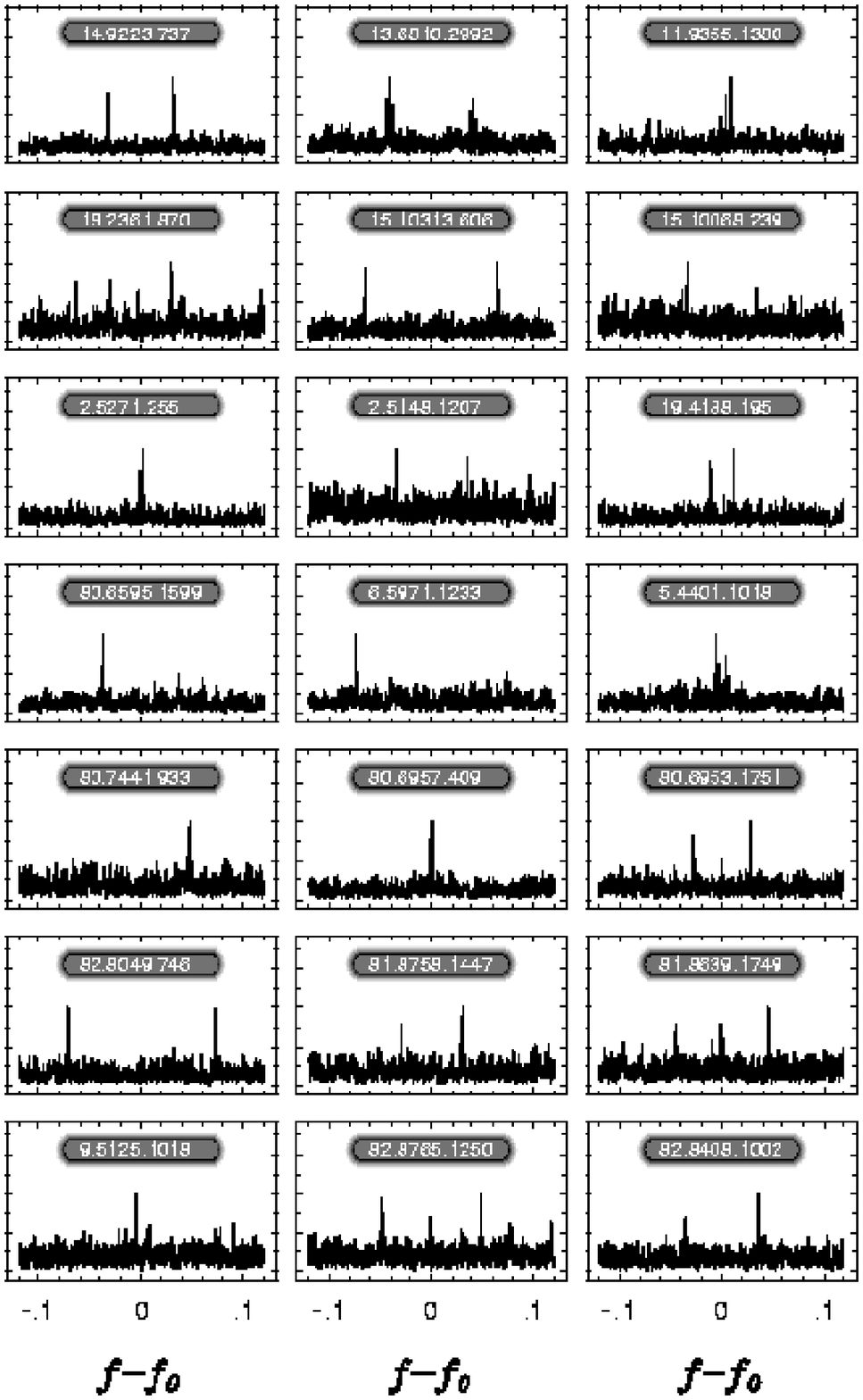,height=140mm,width=110mm}}
\vskip -0mm
\noindent 
{\footnotesize {\bf Fig. 13} -- 
Amplitude spectra of a sample of Blazhko stars for the data 
after the first prewhitening. Each spectrum is normalized to the same 
level by the corresponding highest peak. 
} 
\vskip 6mm
\noindent
To clarify this question, the following tests are performed. First, we 
select high S/N ratio stars from the monoperiodic RR1 variables. 
Stars which satisfy the 
$(0.16A(\nu_0) - \langle A(\nu) \rangle)/\sigma_{A(\nu)} > 5.0$ 
criterion are chosen for the test. Here $A(\nu_0)$ denotes the amplitude 
of the observed main component, other symbols have the same meaning 
as in Eq. (1). This condition is in line with the detection properties 
of the secondary components discussed in Sect. 2 and implies a modest 
or good chance to find any secondary components with amplitudes greater 
than 16\% of the corresponding main components. We note that the number 
of stars selected with this method is a very sensitive function of the 
required discrimination/detection limits. For example, with the above 
criterion we get 216 variables, whereas with a 15\% limit we get only 140. 

In the next step each time series is prewhitened by the main component.  
Then, two symmetric components are added to the residuals with 
$A_{\pm}/A(\nu_0)=0.10$ or, in another test, with $A_{\pm}/A(\nu_0)=0.15$. 
In both cases $\vert \Delta f_{+}\vert = \vert \Delta f_{-}\vert = 0.04$ 
and the phases are arbitrary but equal. The resulting time series is 
analyzed through three sequential prewhitenings in the 
$[\nu_0-0.12,\nu_0+0.12]$~d$^{-1}$ band, and searched for significant 
peaks (both visually and automatically, by finding the most symmetric 
triplets from the four frequencies with the highest amplitudes). 
%
%
%
\vskip 0mm
\centerline{\psfig{figure=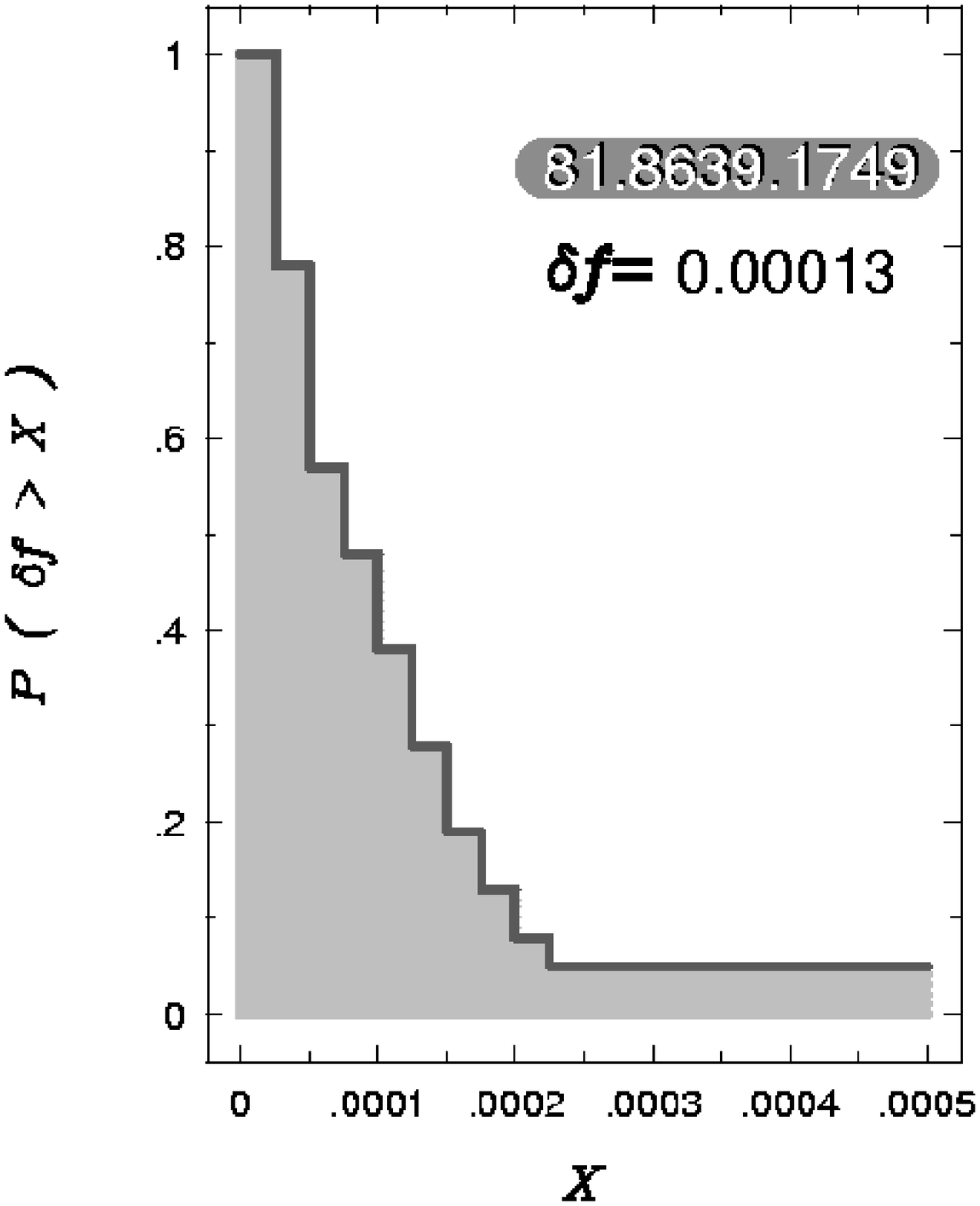,height=100mm,width=90mm}}
\vskip -8mm
\noindent 
{\footnotesize {\bf Fig. 14} -- 
Empirical probability distribution function of the 
deviation from equidistant frequency pattern due to observational 
noise. The observed deviation is given in the top right region of 
the figure under the variable identification code.   
} 
\vskip 8mm
\noindent
By allowing a maximum frequency asymmetry of $\delta f = 0.001$~d$^{-1}$, 
in the $A_{\pm}/A(\nu_0)=0.15$ test we find 155 cases where the two 
symmetric peaks can be identified. In the case of weaker secondary 
components ($A_{\pm}/A(\nu_0)=0.10$), we still have 49 successful 
identifications based on the existence of symmetric triplet structures 
among the four highest peaks obtained during the prewhitening sequences. 
A `conservative' visual examination of the spectra after the first 
prewhitening leads to a similar conclusion. In the test of 10\% 
modulation level we find 49 cases where only one significant component 
is seen (6 of them do not appear at the test frequency). In 22 additional 
cases we see clear equidistant triplet structures. It is evident that if 
we assume a 20\% occurrence rate of these Blazhko stars with low-level 
modulation, in this limited sample of 216 stars alone we should have 
seen more than $0.2\times(49+22) = 14$ of them either as RR1$-\nu1$ 
or RR1$-BL$ variables.

In Figure~15 we display some of the spectra to visualize the significance 
of the detection of the side components in the test case of the 15\% 
modulation level. From the 30 cases shown, there are only two in which 
none of the components are readily visible.  

Considering only the test results of the automatic search for triplet 
structures, let us assume that the incidence of the RR1$-BL$ stars with 
these low-amplitude modulations follows the rate of $\approx 20\%$ found 
for the RR0 stars. Then, with the minimum amplitude modulation of 10\%, 
the conservative estimate is that we expect at least some $49\times 
0.2 \approx 10$ additional RR1$-BL$ stars in the original data set. 
(With the 15\% minimum modulation level this number would increase 
to 31.) To test if we have missed some faint close components during 
the basic screening of 
%
%
\vskip -0mm
\centerline{\psfig{figure=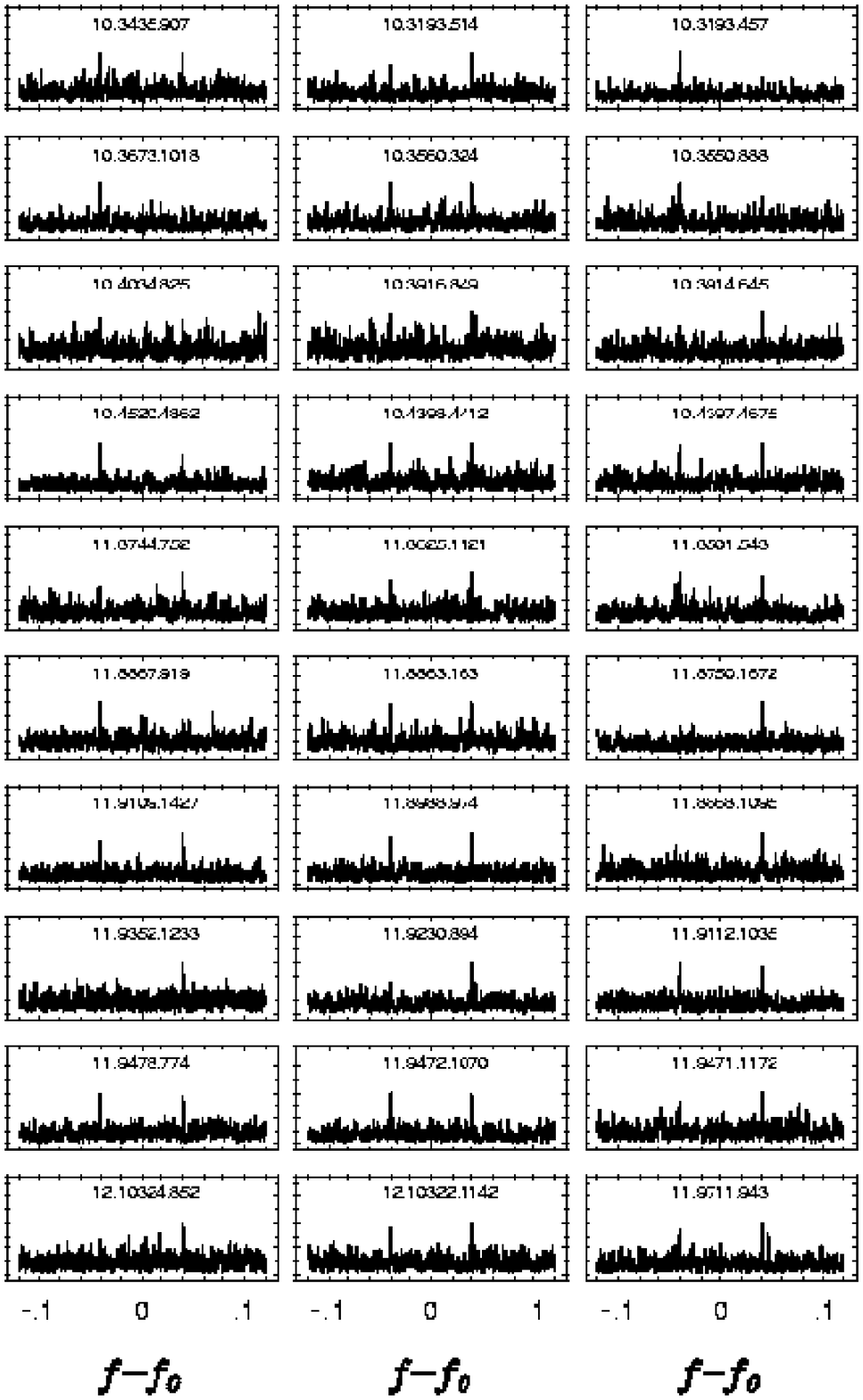,height=140mm,width=110mm}}
\vskip -0mm
\noindent 
{\footnotesize {\bf Fig. 15} -- 
Amplitude spectra of a sample of the {\bf test signals}  
{\it after the first prewhitening}. The {\bf artificial side components} 
correspond to 15\% modulation level. Each spectrum is normalized to 
the same level by the corresponding highest peak. 
} 
\vskip 10mm
\noindent
the data, we repeat the above analysis on the 
original data, without the injection of the faint secondary artificial 
signal components. A visual inspection of the spectra of the 216 stars 
confirms our first conclusion, namely, these stars are indeed 
monoperiodic variables. In Figure~16 we show the 12 best cases, where a 
search for symmetric triplet structure among the highest peaks indicates 
the presence of such hidden triplets. Comparing with the examples of the 
15\% test signal case, we see that even these `best' cases are of very 
low statistical significance. From these tests we conclude that if 
RR1$-BL$ stars with low modulation levels of 10--15\% exist at a rate 
more than 10\%, we should have found such stars in a significant number. 
It seems rather probable that the incidence of the Blazhko stars among the 
LMC RR1 stars cannot be larger than a {\it few percent}.       
%
%
\vskip 7mm
\centerline{\psfig{figure=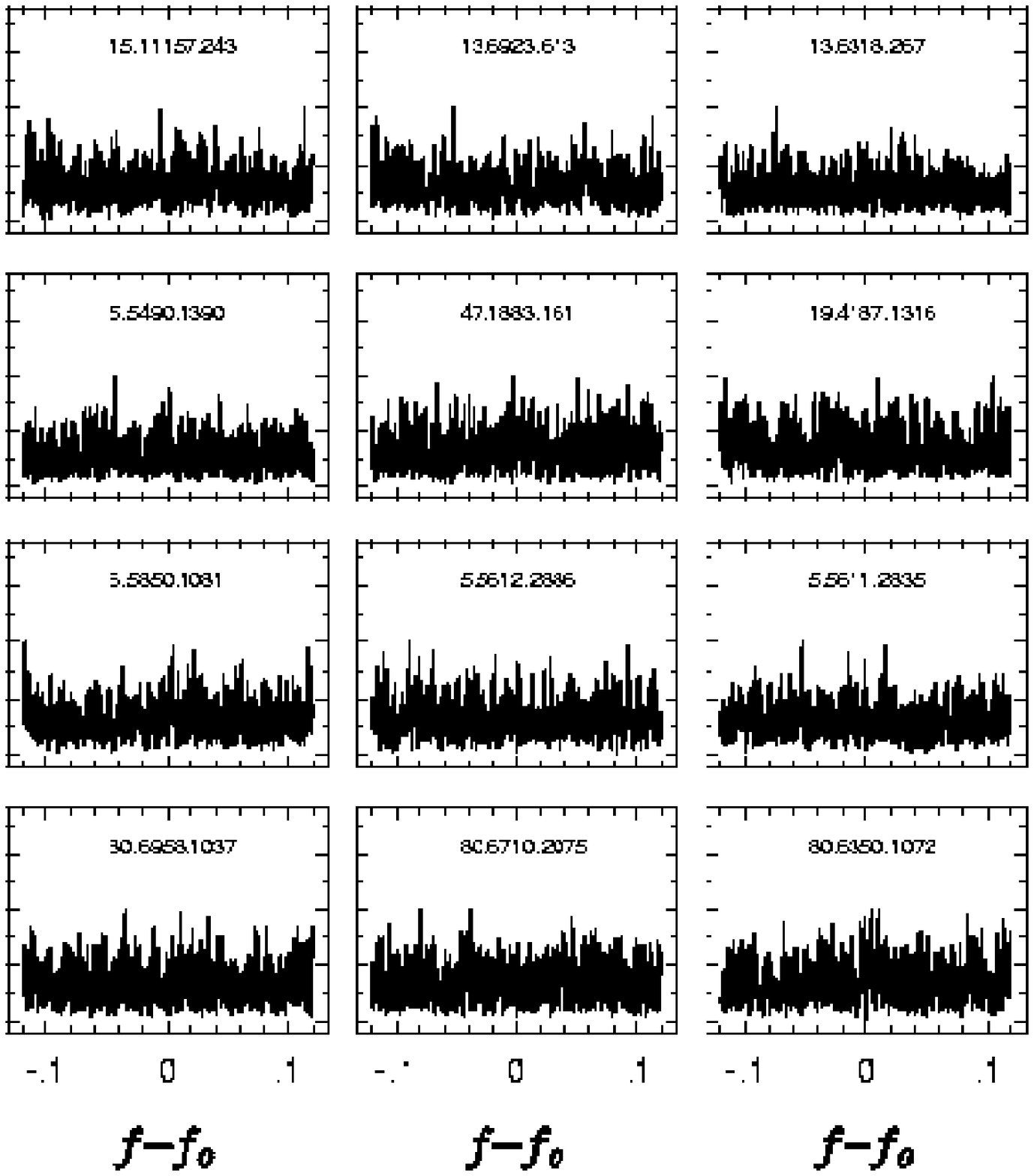,height=90mm,width=100mm}}
\vskip -0mm
\noindent 
{\footnotesize {\bf Fig. 16} -- 
Amplitude spectra of a sample of monoperiodic variables 
{\it after the first prewhitening}. For these variables a numerical 
search for hidden symmetric frequency triplets led to positive results. 
Each spectrum is normalized to the same level by the corresponding 
highest peak.  
} 
\vskip 10mm
%
%
 
%
%
\section{RR1$-PC$-TYPE STARS: VARIABLES WITH PERIOD CHANGES}

In the course of the basic scan of the variables, we found a 
considerable number of stars whose prewhitened frequency spectra  
contained significant remnant power very close to the main component. 
Due to the proximity and unresolved nature of these remnants, we 
attribute their existence to long-term period and/or amplitude 
changes. Quite often the phenomenon can also be recognized by the 
broadening of the line profiles in the frequency spectra. Because 
of the non-discrete nature of the frequency spectra, the standard 
prewhitening procedure does not work in these cases. In Figure~17 we 
show an example of the failure of successive prewhitenings. Spectral 
line broadening is also well observable in the present case.   

For a closer examination of the nature of the long-term variation, 
we perform the following {\it simple time-dependent Fourier analysis}. 
A sequence of single-component Fourier fits is calculated on overlapping 
segments of the time series by shifting the base of the fit by one item 
of data in each step. The length of the base of the fits is chosen to 
ensure reasonable stability in the calculated time-dependent amplitudes 
and phases when the base-length is increased. In most cases we reach this 
situation by choosing 50--100 data points per segments. We also test 
two simple mathematical models of the long-term modulation. The first 
one consists of a linear period change of the main component 
%
\begin{eqnarray}
P(t) = \langle P \rangle + \beta (t-t_0) \hskip 2mm , 
\end{eqnarray}
where $\langle P \rangle$ is a properly chosen average period, $t_0$ 
is an arbitrary epoch and $\beta$ is the rate of period change. We 
expect this model to work well when the phase variation derived above is 
closest to a second order polynomial. The second model assumes a pure 
exponential modulation in the following form 
%
\begin{eqnarray}
A(t) = \langle A \rangle e^{\eta(t-t_0)} \hskip 2mm ,  
\end{eqnarray}
where $\langle A \rangle$ is some average amplitude. 
%
%
\vskip -0mm
\centerline{\psfig{figure=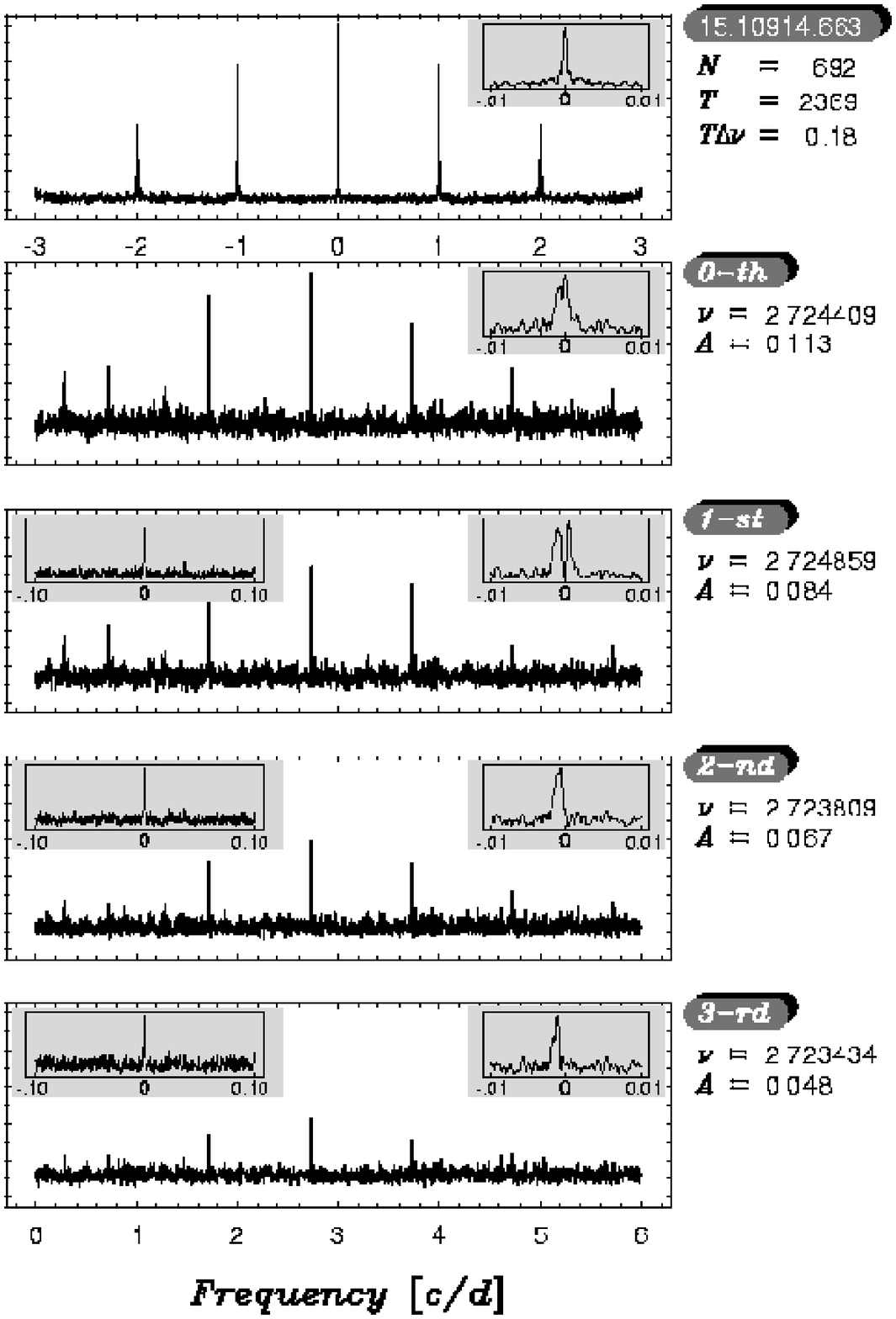,height=130mm,width=90mm}}
\vskip -0mm
\noindent 
{\footnotesize {\bf Fig. 17} -- 
Detection of an RR1$-PC$ variable (RR1 star with 
unresolved components at the main frequency). For notation, see 
Figure~3.
} 
\vskip 3mm
\noindent
After performing these tests on the 141 RR1$-PC$ variables, we find that 
{\it none} of them can be modeled by the assumption of pure exponential 
amplitude change. On the other hand, the model of linear period change 
works better in a significant number of cases. However, the time-dependent 
Fourier analysis reveals that for most of the RR1$-PC$ variables these simple 
models are unable to give a full account of the observed remnant power 
in the frequency spectra. In Figure~18 we show representative cases for 
the almost pure linear period change and for the more general situation 
when this simple description is not applicable. It is important 
to remark that the present simplified time-dependent analysis can be regarded 
as very preliminary, because the high noise level and the strong period 
change would require the application of a more sophisticated method. 
For instance, tiny amplitude changes (if they exist), are certainly not 
detected with this method due to the large errors of the calculated 
amplitudes from the short data segments in the presence of such a high 
noise. However, the observed phase changes seem to have much higher 
significance because of the considerable range of phase variations 
associated with them. In addition, a comparison with a similar 
analysis performed on some RR1 variables reveals that indeed, they 
show irregular/noisy phase variations several times smaller. At the 
same time, their temporal amplitudes behave very similarly to those 
of the RR1$-PC$ stars.    
%
%
\vskip 0mm
\centerline{\psfig{figure=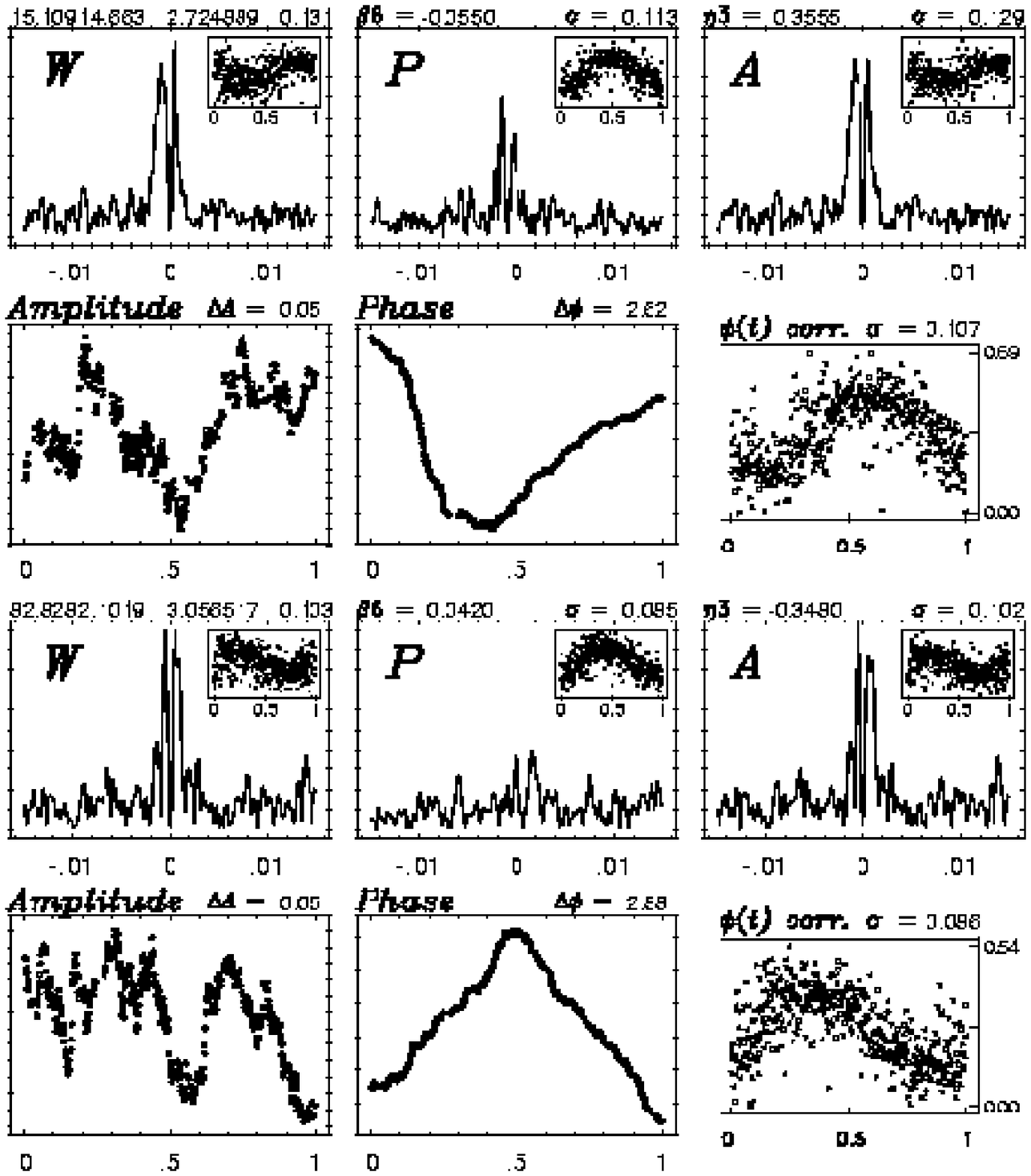,height=90mm,width=100mm}}
\vskip -0mm
\noindent 
{\footnotesize {\bf Fig. 18} -- 
Results of the various tests performed on two 
RR1$-PC$-type variables. {\it Uppermost row:} prewhitened spectra 
calculated in the regime of the main component with the assumptions 
of $P=const.$ (W), linear period change (Eq. (3), P) and exponential 
amplitude change (Eq. (4), A). All spectra are normalized by the 
highest peak of the W spectrum. {\it Insets:} corresponding 
folded light curves. {\it Headers:} star code, main frequency 
[d$^{-1}$], $\sigma$ [`r' mag] of the folded light curve, 
$\beta 6 = \beta \times 10^6$~[dd$^{-1}$], 
$\eta 3 = \eta \times 10^3$~[d$^{-1}$], the sigmas refer to the folded 
light curves obtained by the corresponding assumptions. 
{\it Next row:} time-dependent amplitude, phase and folded light curve 
corrected by the plotted phase variation. The horizontal axes in the 
first two figures represent the total time span. {\it Headers:} 
the total amplitude and phase ranges shown, $\sigma$ of the folded 
light curve. {\it Subsequent rows:} as above, but for a different 
variable.   
} 
\vskip 6mm
\noindent
We draw the attention to the fact already mentioned, namely that a 
simple exponential amplitude variation cannot explain the remnant 
power in the spectrum. Furthermore, the light curves corrected for 
the phase variation show considerably lower scatter. For these reasons 
we refer to the above class of variables as $PC$ (period changing) 
stars, since most, if not all their extra power around their main 
frequency components originates from phase/frequency change, although 
some contribution from amplitude change cannot be excluded. 

If we consider the signs and sizes of the derived period change rates, 
we see that: (a) there are both positive and negative values; (b) in 
absolute values they are about two orders of magnitude larger than the 
ones expected from standard stellar evolution theory (e.g., Lee 1991). 
This type of discrepancy (although perhaps at a lower degree) exists 
for most of the observed period changes in RR~Lyrae stars. The 
explanation of this fact poses a serious challenge for stellar pulsation 
theory. Because the individual period changes are well-demonstrated and 
rather accurately measured, we think that it is incorrect to treat large 
period changes as errors and use only their averages over the variables 
studied (e.g., Lee 1991; Lee \& Carney 1999; see however Rathbun \& Smith 
1997). 

We list the RR1$-PC$ variables found during this survey in Table 4. 
We caution that in some cases three prewhitenings with the main and 
two very close side components yield flat flat frequency spectra 
(e.g., 3.6240.470; 6.6568.484; 11.9235.1034). This suggest that these 
stars might turn out to be RR1$-BL$ variables with very long modulation 
periods when future observations on a longer time base will be available.      

%
%
\section{MISCELLANEOUS VARIABLES}

Either because of their rare occurrence or because of the specific 
frequency values of their secondary components, it is difficult to 
classify these stars with high confidence. Nevertheless, future 
studies might shed light both on the nature of these stars and on 
the possible relations between them and the more securely classified 
variables. 
%
%
\vskip -0mm
\centerline{\psfig{figure=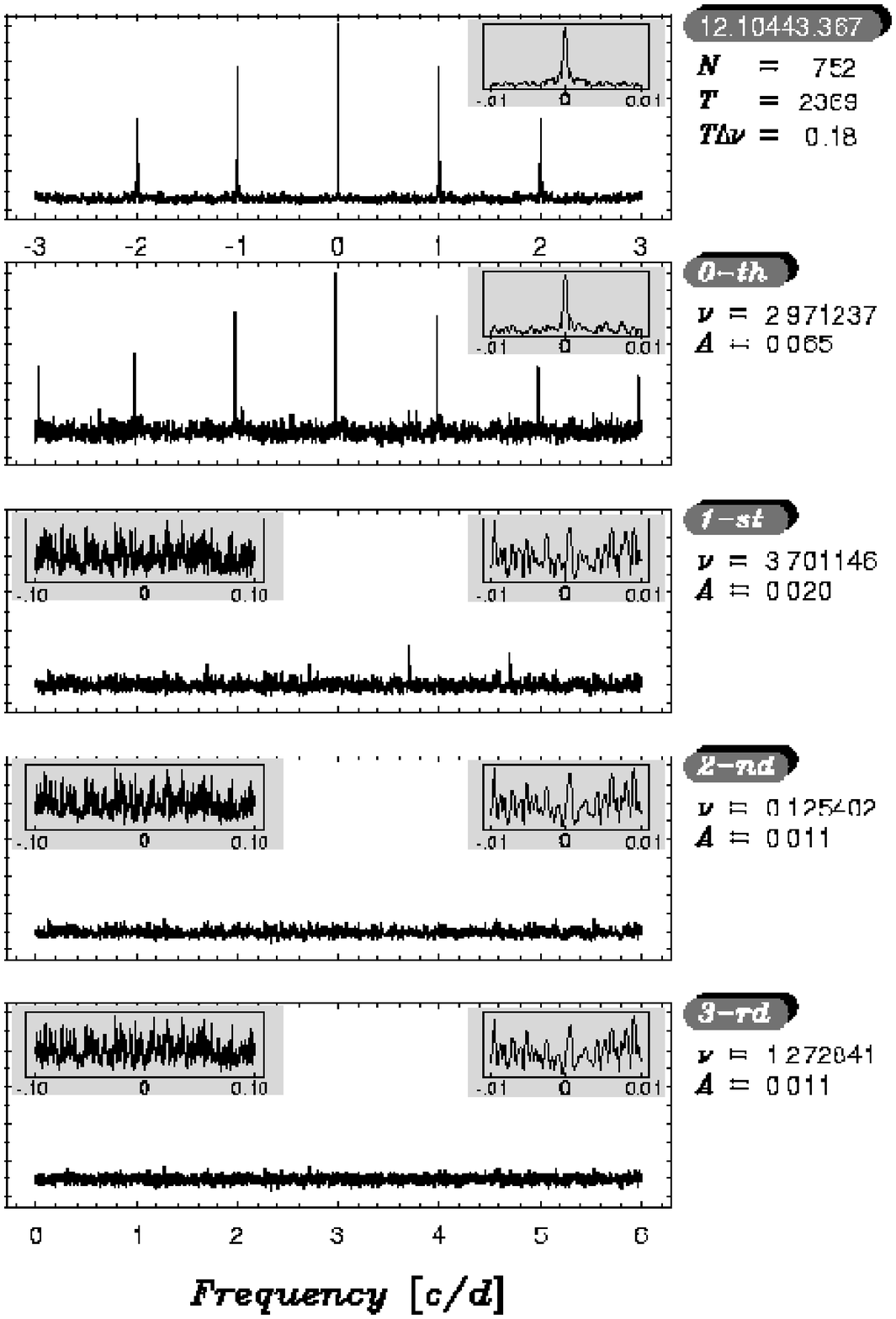,height=130mm,width=90mm}}
\vskip -0mm
\noindent 
{\footnotesize {\bf Fig. 19} -- 
A suspected double-mode variable pulsating in the first 
and second overtones. For notation, see Figure~3.
} 
\vskip 3mm
\noindent
There are 20 stars whose prewhitened frequency spectra contain residuals 
at integer d$^{-1}$ components. Hereafter we call these variables of 
RR1$-D1$-type. Usually, the amplitudes of these curious components are not 
too large, exceeding 50\% of the main components only in two variables. 
For half of them this quantity is under 30\%. In general, the secondary 
components show up at 1.00027~d$^{-1}$ but sometimes the $\pm 1$~d$^{-1}$ 
aliases are stronger. As a result, it is possible that in some cases 
a long-term effect is responsible for the observed frequency pattern. 
We find many cases when the `b' data do not justify the RR1$-D1$ 
classification. For these variables we retain their RR1 status. 
Most probably, the RR1$-D1$ variables are affected by some spurious 
observational or data reduction flaws, to be clarified in the future. 

In three more interesting cases the frequency ratios of the secondary 
and main components closely resemble those of the first and second 
overtone modes of RR~Lyrae models. Therefore, these preliminarily 
classified double-mode variables are referred to as RR12-type. One of 
them is shown in Figure~19. Further studies of these variables would be 
important for the verification of their double-mode behavior and their 
relations to the single-mode second overtone suspects (Alcock et al. 1996). 
%
%
\vskip -0mm
\centerline{\psfig{figure=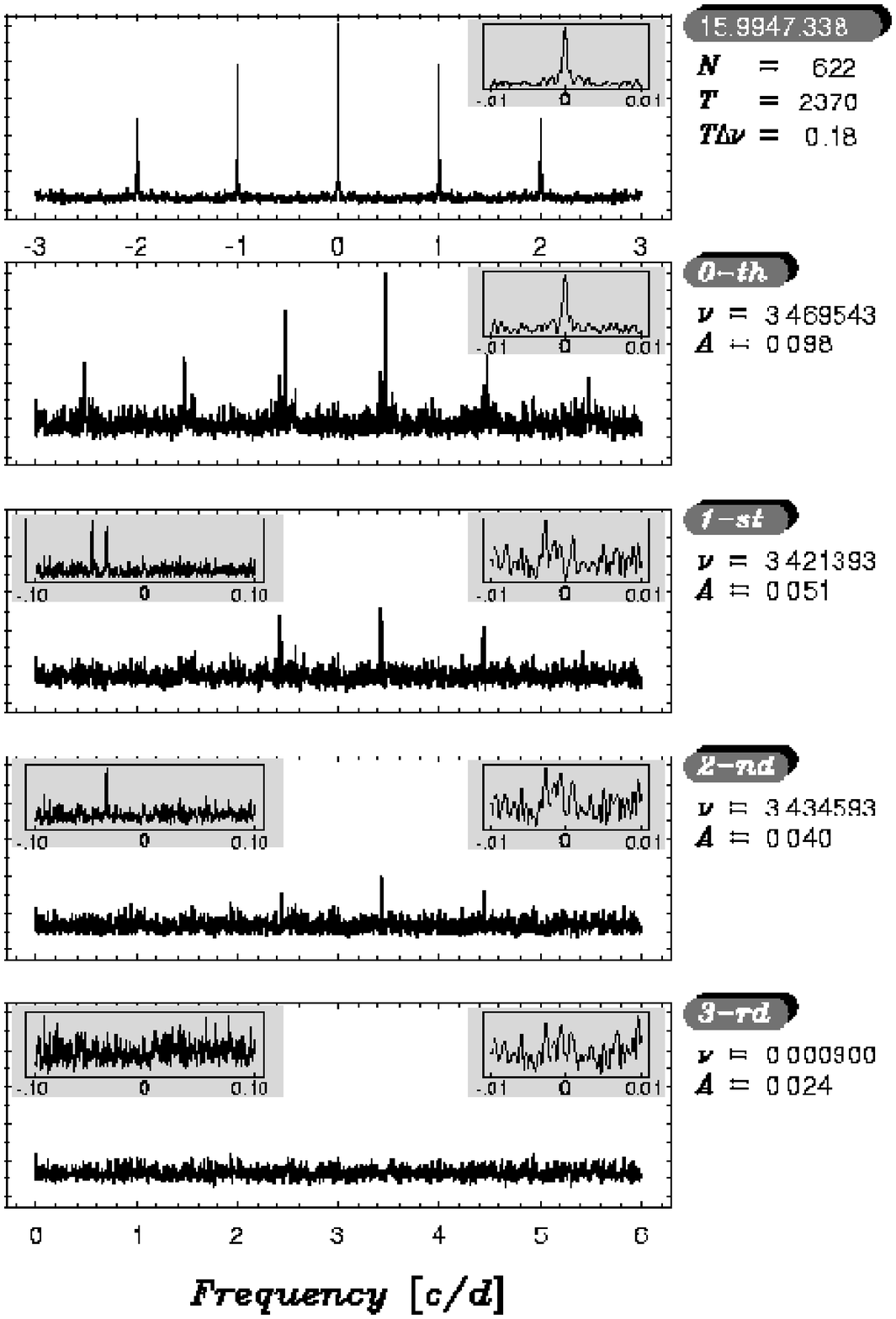,height=130mm,width=90mm}}
\vskip -0mm
\noindent 
{\footnotesize {\bf Fig. 20} -- 
An example of the few RR1$-\nu2$-type variable found in the 
RR1 sample. For notation, see Figure~3.
} 
\vskip 3mm
\noindent
In the frequency spectra of three variables we find two significant 
peaks close to the main component. However, unlike in the case of Blazhko 
stars, these peaks are {\it not spaced equidistantly}. We refer to 
these stars as RR1$-\nu2$-type variables. Figure~20 displays the interesting 
case of 15.9947.338, where the two additional components are located 
close to each other, but further away from the main component. Another 
star, 2.5023.5787 shows three peaks of equal heights and asymmetric 
frequency spacing ($\delta f = 0.0030$~d$^{-1}$). Because of the proximity 
of their frequency components, one may wonder whether the RR1$-\nu1$, 
RR1$-\nu2$- and Blazhko-type phenomena are physically related. At this 
moment it is not possible to deal with this question. Further 
detailed observational and theoretical works are needed. 
%
%
\vskip -0mm
\centerline{\psfig{figure=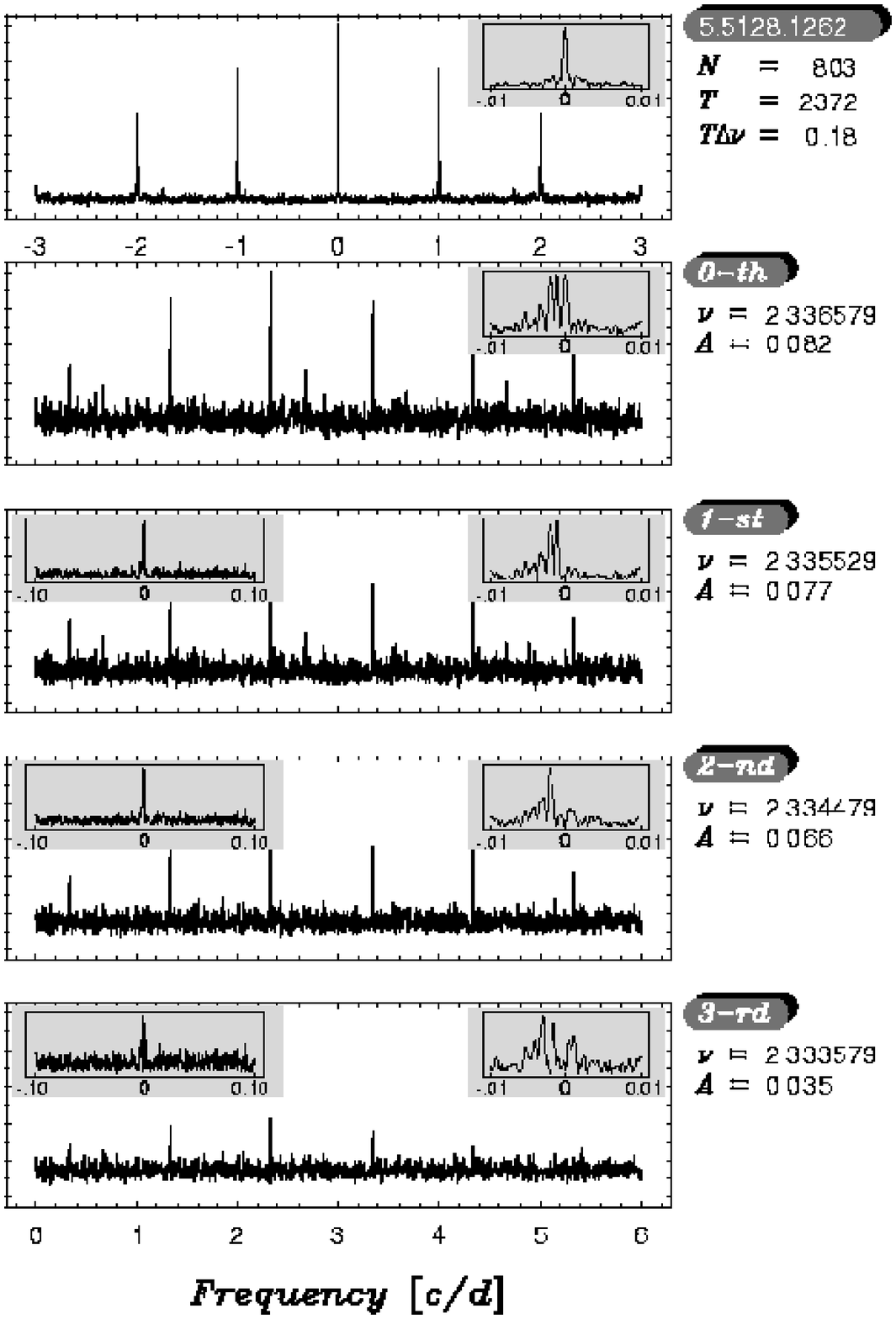,height=130mm,width=90mm}}
\vskip -0mm
\noindent 
{\footnotesize {\bf Fig. 21} -- 
An example of the few RR1$-\nu M$-type variable found in the 
RR1 sample. For notation, see Figure~3.
} 
\vskip 3mm
\noindent
In five cases there are hints that the prewhitened spectra consist of 
several well-resolved peaks located close to the main component. Figure~21 
displays the frequency spectra of one such variable. Following the 
nomenclature of the other variables with close components, we refer 
to these stars as RR1$-\nu M$-type variables. It is tempting to identify 
these objects with multimode variables whose secondary frequency 
components correspond to some of the most strongly excited nonradial 
modes belonging to the same harmonic degree $l$ (for recent theoretical 
models see Dziembowski, \& Cassisi 1999). However, it is better 
to wait with this conclusion until more elaborated observational 
tests become available. 

We also mention the relatively small group of variables whose 
classification is hampered by the complexity of their spectra. 
This complexity is either due to the appearance of peaks without 
any hints of their origin, or to the high noise/low data number. 
These non-classified (RR1$-NC$) variables are listed together with 
Figure~e above mentioned miscellaneous objects in Table 5. For further 
reference, in Table 6 we give the frequencies and amplitude ratios 
of the RR12 and RR1$-\nu2$ stars. The symbols $\nu_i$ refer to the 
frequencies obtained at the various stages of the prewhitening sequences, 
and not to the normal modes, within the system of standard notation 
for these modes.   

It is important to recall that in some cases, in other 
types of variables described in the previous sections, we also 
encountered some peculiarities. First of all, because of the 
finite noise level and of the color dependence of the observed 
amplitudes, the `r' and `b' data sometimes yield inconsistent 
results. These situations occur usually when the secondary 
component is close to the noise level. In these cases classification 
is based on the 'cleanest' frequency spectrum. For RR1$-D1$-type 
variables this practice is partially overruled by the fact that the 
secondary component is most probably an artifact, therefore, preference 
is given to the monoperiodic RR1 classification if it is supported by 
one of the colors. As another example for the possible ambiguity, 
we recall the case of some RR1$-PC$ variables mentioned in Sect. 6. 
In the case of these stars the length of the available data is still 
too short to enable us to make a clear distinction between period 
change and the more than $\approx 2000$~d modulation period if the 
variable is of Blazhko-type, as it is suggested by the successive 
prewhitenings. 
 
%
%
\section{CONCLUSIONS}
   
A massive frequency analysis has been carried out on a large sample 
of the first overtone RR~Lyrae (RR1) population of the {\sc macho} 
variable star base of the Large Magellanic Cloud. The study was aimed 
primarily at finding  multiperiodic and non-stationary variables among 
RR1 stars. Special effort was made to search for amplitude- and 
phase-modulated (Blazhko) variables. Occurrence of this phenomenon 
among RR1 stars was not known previously. Because widely acceptable 
theoretical interpretation of the Blazhko effect is still missing, 
it is important to accumulate more observational data in order to 
constrain further the possible models. Because of the high occurrence 
rate of the Blazhko effect among fundamental mode RR~Lyrae stars and 
the lack of theoretical understanding of this phenomenon, known for 
almost a century, further study of these stars has a general significance. 
%
%
\begin{deluxetable}{llrrc}
\tablecaption{Variable types in the {\sc macho} database for the 
first overtone RR~Lyrae stars in the LMC }
\tablenum{7}
\tablehead{
{\it Type} & {\it Short description} & {\it Number} & \% 
           & $\sigma$(\%) }
\startdata
RR1         & Single-mode                     & 916 & 69.0 & 1.3 \\
RR1$-\nu1$  & 1 close component               &  24 &  1.8 & 0.4 \\
RR1$-\nu2$  & 2 asymmetric close components   &   3 &  0.2 & 0.1 \\
RR1$-\nu M$ & Multifrequency close components &   5 &  0.4 & 0.2 \\
RR1$-BL$    & 2 symmetric close components    &  28 &  2.1 & 0.4 \\
RR1$-PC$    & Period change                   & 141 & 10.6 & 0.9 \\
RR01        & 1st \& 0th overtone double-mode & 181 & 13.6 & 0.9 \\
RR12        & 2nd \& 1st overtone double-mode &   3 &  0.2 & 0.1 \\
RR1$-D1$    & Integer d$^{-1}$ frequencies    &  20 &  1.5 & 0.3 \\
RR1$-NC$    & Non-classified                  &   6 &  0.5 & 0.2 \\
\enddata
\tablecomments{The symbol $\sigma$(\%) denotes the standard 
deviation of the population ratio, assuming Poisson distribution 
in the different populations.} 
\end{deluxetable}

The basic technique of analysis we applied was the standard Fourier 
frequency analysis with the important sequential steps of prewhitening. 
All 1350 variables passed through a basic analysis which resulted in the 
selection of more than 400 multimode candidates for further studies. 
In Table 7 we summarize the statistics of the various types of variables 
we identified in the sample. This table does not contain the more than 
30 doubly identified variables which were among the 1350 stars in the 
basic analysis of the data set. However, to make the statistics more 
complete, we added the 73 double-mode (RR01) variables of Alcock et al. 
(1997) to the newly discovered 108 stars of this study. As we have 
already warned in the paper, the classification of some of the variables 
may not be definitive. Although this results in some ambiguity in the 
derived statistics, it does not change our basic conclusions. Furthermore, 
because of the finite noise level, there is a limit for the lowest 
detectable secondary signal components. Considering specifically 
the Blazhko-type stars, from the statistical tests presented in 
Sect. 5, we estimate this limit to be 10--15\% in the units of the 
main signal component (10\% being the level of marginal, whereas 15\% 
is the level of close to secure detectability).   

The 108 newly discovered RR01 stars follow the pattern on the 
$P_0\rightarrow P_1/P_0$ diagram of the already known 73 double-mode 
variables (Alcock et al. 1997). This pattern implies metallicity and/or 
mass spread among the RR01 population of the LMC. Direct metallicity 
measurements through the $\Delta S$ method by Clementini et al. (2000) 
suggest the existence of sufficient [Fe/H] spread. The preliminary 
theoretical work of Popielski \& Dziembowski (2000) comes to a similar 
conclusion. Accurate photometry and metallicity measurements of these 
stars would be very valuable also for constraining further the LMC 
distance scale (Kov\'acs \& Walker 1999). 

We found three variables suspected for double-mode pulsations in the 
first and second overtones. Further study of these stars with additional 
photometry would be necessary to confirm their suggested pulsational 
status. 

A considerable fraction of the RR1 population shows period change. 
Although some of them can be approximated by linear period changes, 
their high values exclude any evolution-related explanation. Understanding 
period changes in these and in other RR~Lyrae stars remains an important 
unsolved problem in stellar pulsation theory. 

We regard the most exciting finding of this study the discovery of the 
variables with closely spaced frequencies (for preliminary works on these 
and some fundamental mode Blazhko stars we refer to Kov\'acs et al. 2000 
and Kurtz et al. 2000). About half of these stars show only two components, 
similarly to the recent discoveries of the same type of stars by 
Olech et al. (1999a,b) and Moskalik (2000). Although in some of these 
RR1$-\nu1$-type stars in our sample a weak third component with 
symmetric frequency spacing can be suspected, neither the large majority 
of our RR1$-\nu1$ stars, nor the variables reported by other authors 
show such components. However, in 28 variables we observe symmetric 
frequency spacing, very similar to the frequency spectra of the 
fundamental mode (RR0) Blazhko-type variables (e.g., Kov\'acs 1995). 
Therefore, we associate these variables with the Blazhko-type stars. 
At this moment it is only the LMC in which first overtone Blazhko stars 
have been identified. 

Compared with the commonly used incidence rate of 20--30\% of the RR0 
Blazhko variables (Szeidl 1988), our 2\% frequency among RR1 stars seems to 
indicate that whatever is the underlying cause of the Blazhko phenomenon, 
it should work quite {\it differently} in RR1 than in RR0 stars. Although 
in a milder sense, this difference still survives even if we consider the 
recent statistics obtained by Moskalik (2000) on the {\sc ogle} sample 
of Galactic bulge RR~Lyrae stars. He found 11 Blazhko variables from the 
149 RR0 stars but none from the 66 RR1 stars analyzed (however, he 
found RR1$-\nu1$-type stars: 11 among the RR0, and 2 among the RR1 stars). 
According to our tests, if the lowest level of modulation among RR1 stars 
were 15\%, assuming 20\% incidence rate for the Blazhko phenomenon,  
we should have detected at least 30 more variables of this type. Although 
with a 10\% modulation level under the same circumstance this number 
decreases to about 10, it is still large enough to conclude that 
we probably have not missed any Blazhko RR1 variables above the 10\% 
modulation level. Adding to these that the high frequency of the Blazhko 
RR0 stars refers to relatively older data with marginal probability 
of the detection of low-amplitude modulations, and that most of them 
have 30--50\% total modulation level, we think that the large difference 
in the occurrence rates of the two types of Blazhko stars is significant. 

It follows from this result that {\it magnetic oblique rotator} model 
(Shibahashi 1994; 2000) of the Blazhko phenomenon must face with the 
difficulty in explaining why the role of magnetic field becomes so 
much less important as a star pulsating in the fundamental mode and 
showing Blazhko phenomenon moves to the RR1 regime. Most of these 
stars must either stop being Blazhko-type variables, or their modulation 
level must decrease from 30--50\% down to below 10\% (in the units of 
the total pulsation amplitude). In addition, there is also the problem 
of asymmetric modulation amplitudes. This difficulty becomes extreme for 
the above theory to explain both Blazhko- and RR1$-\nu1$-type variables 
within the same framework. In order to deal with these basic observational 
facts, the magnetic model needs to be more complex than it is now. 
This, without the precise measurement of magnetic field, would lead 
to introducing more free parameters, and thereby to decreasing the 
predictive power of the theory. 
      
The other theory, the model of {\it direct ($\omega_0\approx \omega_l$) 
resonant mode coupling between a radial and nonradial modes} stays 
within the framework of nonlinear stellar pulsation, without invoking 
the presence of magnetic field. According to Van Hoolst, Dziembowski 
\& Kawaler (1998), with significant probability this resonance is 
capable of destabilizing the fundamental limit cycle and leading either 
to steady {\it phase locked} pulsation, or to an amplitude- and 
phase-modulated one. In both cases we would see Blazhko variability 
but in the steady amplitude and phase solution observability would 
depend on the aspect angle of the observer relative to the rotation 
axis. Concerning the resonance theory, it is important to remark that 
the recent linear pulsation survey of Dziembowski \& Cassisi (1999) shows 
that the excitation rates for the resonant $l=1$ modes are comparable 
to those of the radial fundamental and first overtone modes. Although 
these results are encouraging, many questions remain to be answered, 
such as the relation between RR1$-\nu1$- and Blazhko-type stars, and, 
of course, the low number of the Blazhko RR1 stars. 

We look forward to completing a similar survey of multimode RR~Lyrae 
stars in our six fields in the Small Magellanic Cloud. While we do not 
anticipate significant differences in the fractions of the types of 
multimode RR~Lyrae variables or sample metallicity, such a survey will 
provide a test of the degree to which such samples are homogeneous. 
In many cases the signal-to-noise of the SMC RR~Lyrae photometry is 
expected to be better than that of the LMC photometry because our 
SMC exposure times were twice as long and this more than compensates 
for the greater distance of the SMC.

%
%

\acknowledgements
This work was initiated during GK's stay at the IGPP of the Lawrence 
Livermore Laboratory. He thanks the director and the staff for their 
hospitality. Fruitful discussions with Hiromoto Shibahashi are very 
much appreciated. We thank Wojtek Dziembowski for his valuable comments
as a referee.  
We are very grateful for the skilled support given our project 
by the technical staff at the Mt. Stromlo Observatory.  
Work performed at LLNL is supported by the DOE under contract 
W7405-ENG-48.
Work performed by the Center for Particle Astrophysics on the UC campuses
is supported in part by the Office of Science and Technology Centers of
NSF under cooperative agreement AST-8809616.
Work performed at MSSSO is supported by the Bilateral Science 
and Technology Program of the Australian Department of Industry, 
Technology
and Regional Development. KG acknowledges a DOE OJI grant, and CWS
and KG thank the Sloan Foundation for their support.
DLW was a Natural Sciences and Engineering Research Council (NSERC)
University Research Fellow during this work.
DM is supported by Fondecyt 1990440. 
The supports of the following grants are acknowledged: 
{\sc otka t$-$024022, t$-$026031} and {\sc t$-$030954}.


%
%

\newpage

%
%

{\tiny
\begin{deluxetable}{lcccccc}
\tablecaption{Previously not known RR01 variables of the MACHO RR1 sample}
\tablenum{1}
\tablehead{
{\it MACHO \#}& $\alpha$ & $\delta$& $P_1$ & $P_1/P_0$& $A_1$ & $A_0/A_1$ }
\startdata
80.7193.1485   &05:24:39.0& -69:20:27 & 0.328818 & 0.7419 & 0.163 & 0.319 \\
81.8639.1450   &05:33:27.0& -69:42:19 & 0.335257 & 0.7415 & 0.091 & 0.396 \\
10.4043.1392   &05:05:43.3& -69:34:13 & 0.343077 & 0.7427 & 0.128 & 1.078 \\
80.7439.1836   &05:26:14.5& -69:03:33 & 0.343358 & 0.7429 & 0.136 & 1.184 \\
5.4889.1102    &05:10:35.6& -69:39:11 & 0.343666 & 0.7427 & 0.108 & 1.009 \\
14.9220.799    &05:37:10.7& -71:21:15 & 0.343895 & 0.7433 & 0.152 & 0.704 \\
5.5254.1692    &05:12:57.8& -69:33:39 & 0.345360 & 0.7430 & 0.112 & 0.384 \\
5.4524.1212    &05:08:13.5& -69:46:46 & 0.346200 & 0.7435 & 0.152 & 0.322 \\
14.9467.864    &05:38:53.0& -71:00:13 & 0.346562 & 0.7437 & 0.116 & 0.362 \\
80.7073.1658   &05:24:00.3& -69:15:00 & 0.347832 & 0.7433 & 0.141 & 0.823 \\
6.6086.792     &05:17:59.1& -70:30:25 & 0.348080 & 0.7438 & 0.128 & 0.344 \\
82.8410.986    &05:32:28.1& -68:52:04 & 0.348386 & 0.7432 & 0.142 & 0.563 \\
3.7448.428     &05:26:20.0& -68:28:54 & 0.348508 & 0.7429 & 0.123 & 0.382 \\
15.10797.871   &05:46:46.9& -71:03:52 & 0.348567 & 0.7430 & 0.134 & 0.701 \\
3.7447.739     &05:25:54.5& -68:29:42 & 0.349064 & 0.7437 & 0.126 & 0.286 \\
81.8882.1067   &05:34:51.2& -69:41:06 & 0.349122 & 0.7431 & 0.101 & 0.851 \\
6.6214.4637    &05:18:57.3& -70:02:04 & 0.349177 & 0.7432 & 0.086 & 0.930 \\
81.9118.1794   &05:36:42.8& -70:05:01 & 0.350143 & 0.7432 & 0.148 & 0.939 \\
2.4904.1651    &05:10:20.5& -68:39:40 & 0.350355 & 0.7437 & 0.126 & 0.635 \\
9.5487.736     &05:13:58.4& -70:09:06 & 0.350612 & 0.7436 & 0.123 & 0.447 \\
18.2717.812    &04:56:51.9& -69:16:01 & 0.350658 & 0.7440 & 0.116 & 0.586 \\
80.6590.1844   &05:21:07.3& -69:11:15 & 0.351176 & 0.7437 & 0.122 & 0.615 \\
9.4873.519     &05:10:50.7& -70:42:24 & 0.351258 & 0.7438 & 0.127 & 0.843 \\
14.9587.828    &05:39:21.7& -71:01:47 & 0.351263 & 0.7433 & 0.134 & 0.567 \\
15.10071.888   &05:42:25.3& -71:02:45 & 0.352199 & 0.7439 & 0.116 & 0.957 \\
14.9703.450    &05:40:11.9& -71:23:48 & 0.352214 & 0.7440 & 0.129 & 0.271 \\
11.9838.1201   &05:40:46.4& -70:26:21 & 0.352312 & 0.7442 & 0.149 & 0.215 \\
6.5732.3906    &05:15:56.2& -69:57:23 & 0.352354 & 0.7437 & 0.097 & 0.907 \\
15.10071.765   &05:42:44.5& -71:04:15 & 0.352895 & 0.7442 & 0.138 & 0.558 \\
9.4275.536     &05:07:13.3& -70:17:38 & 0.353072 & 0.7437 & 0.131 & 0.832 \\
9.5124.1174    &05:11:43.8& -70:08:09 & 0.353582 & 0.7439 & 0.148 & 0.669 \\
18.2234.1176   &04:54:36.2& -69:13:03 & 0.353711 & 0.7440 & 0.152 & 0.658 \\
82.8169.654    &05:30:44.5& -68:47:50 & 0.353998 & 0.7442 & 0.112 & 0.509 \\
6.6452.704     &05:20:00.6& -70:20:44 & 0.354096 & 0.7438 & 0.135 & 0.800 \\
2.5274.1185    &05:12:28.3& -68:09:49 & 0.355067 & 0.7438 & 0.129 & 0.465 \\
6.5730.3869    &05:15:48.4& -70:02:35 & 0.355162 & 0.7442 & 0.103 & 0.456 \\
6.5853.3986    &05:16:26.5& -69:57:38 & 0.355413 & 0.7436 & 0.141 & 0.248 \\
3.7331.467     &05:25:43.5& -68:11:51 & 0.355927 & 0.7441 & 0.137 & 0.723 \\
19.3581.468    &05:02:32.6& -68:08:06 & 0.356444 & 0.7440 & 0.140 & 0.750 \\
9.5119.644     &05:12:09.8& -70:28:05 & 0.356593 & 0.7437 & 0.123 & 0.626 \\
13.6925.627    &05:23:35.0& -71:03:46 & 0.356680 & 0.7442 & 0.125 & 0.440 \\
3.6725.519     &05:22:02.1& -68:16:24 & 0.356781 & 0.7441 & 0.139 & 0.410 \\
82.8886.1146   &05:35:21.4& -69:23:09 & 0.357001 & 0.7445 & 0.142 & 0.577 \\
80.7202.4678   &05:24:29.6& -68:45:33 & 0.357115 & 0.7442 & 0.121 & 0.430 \\
80.7071.5289   &05:23:52.4& -69:25:22 & 0.357209 & 0.7441 & 0.162 & 0.673 \\
13.6197.522    &05:18:58.1& -71:11:27 & 0.357237 & 0.7442 & 0.120 & 0.242 \\
3.7082.957     &05:23:52.3& -68:41:00 & 0.357462 & 0.7436 & 0.113 & 0.664 \\
9.4514.979     &05:08:30.1& -70:27:36 & 0.357478 & 0.7446 & 0.144 & 0.257 \\
11.8750.1694   &05:34:14.2& -70:25:18 & 0.357774 & 0.7438 & 0.139 & 0.856 \\
3.6963.574     &05:23:34.4& -68:30:28 & 0.358068 & 0.7447 & 0.131 & 0.603 \\
19.4429.703    &05:07:45.6& -68:03:35 & 0.358129 & 0.7446 & 0.137 & 0.350 \\
5.4892.3500    &05:10:54.2& -69:28:24 & 0.358174 & 0.7444 & 0.165 & 0.382 \\
19.4308.880    &05:06:45.6& -68:03:25 & 0.358336 & 0.7444 & 0.132 & 0.409 \\
6.6574.1353    &05:20:45.0& -70:14:38 & 0.358535 & 0.7444 & 0.149 & 0.289 \\
19.3702.545    &05:03:24.2& -68:08:53 & 0.358756 & 0.7436 & 0.068 & 0.603 \\
19.4057.1115   &05:05:39.6& -68:40:34 & 0.358970 & 0.7443 & 0.130 & 0.231 \\
10.4522.4058   &05:08:25.8& -69:56:21 & 0.358983 & 0.7446 & 0.129 & 0.589 \\
6.6089.1661    &05:17:44.5& -70:19:27 & 0.359199 & 0.7442 & 0.151 & 0.662 \\
10.4035.1299   &05:05:11.2& -70:08:25 & 0.359503 & 0.7440 & 0.132 & 0.705 \\
6.6329.884     &05:19:16.2& -70:26:40 & 0.359788 & 0.7431 & 0.134 & 0.269 \\
82.8889.452    &05:35:01.6& -69:13:24 & 0.360186 & 0.7445 & 0.073 & 0.658 \\
6.6212.1063    &05:18:32.7& -70:12:21 & 0.360280 & 0.7445 & 0.105 & 0.514 \\
13.5842.2316   &05:16:27.6& -70:38:57 & 0.360422 & 0.7451 & 0.140 & 0.664 \\
80.7072.2154   &05:24:05.6& -69:21:13 & 0.360853 & 0.7443 & 0.165 & 0.400 \\
11.9595.1162   &05:39:39.1& -70:30:27 & 0.361067 & 0.7429 & 0.090 & 1.056 \\
19.4541.1357   &05:08:42.6& -68:40:30 & 0.361879 & 0.7436 & 0.128 & 0.938 \\
11.8985.882    &05:35:44.3& -70:53:11 & 0.362499 & 0.7443 & 0.132 & 0.765 \\
5.4767.962     &05:09:56.7& -69:42:31 & 0.363025 & 0.7439 & 0.139 & 0.331 \\
13.6927.606    &05:23:22.7& -70:57:02 & 0.363183 & 0.7450 & 0.133 & 0.195 \\
82.7922.520    &05:29:13.8& -69:08:28 & 0.363555 & 0.7446 & 0.106 & 0.736 \\
12.10682.793   &05:46:27.1& -70:39:40 & 0.363690 & 0.7441 & 0.110 & 0.373 \\
2.5389.1138    &05:13:38.8& -68:36:13 & 0.364125 & 0.7452 & 0.124 & 0.444 \\
15.10557.2954  &05:45:15.9& -70:54:36 & 0.364422 & 0.7444 & 0.130 & 0.623 \\
3.7208.477     &05:24:24.6& -68:21:09 & 0.364541 & 0.7451 & 0.139 & 0.504 \\
9.5238.804     &05:13:02.4& -70:36:07 & 0.365550 & 0.7450 & 0.150 & 0.300 \\
47.2619.1486   &04:56:32.8& -67:44:22 & 0.365873 & 0.7443 & 0.102 & 0.912 \\
9.4881.635     &05:10:52.4& -70:12:50 & 0.366304 & 0.7446 & 0.104 & 0.673 \\
12.11172.776   &05:49:17.8& -70:16:26 & 0.366906 & 0.7451 & 0.118 & 0.636 \\
9.5483.1857    &05:14:21.3& -70:21:47 & 0.367222 & 0.7444 & 0.160 & 0.569 \\
10.3798.927    &05:04:07.1& -69:47:41 & 0.367266 & 0.7443 & 0.102 & 1.000 \\
12.10679.968   &05:45:56.5& -70:50:38 & 0.367586 & 0.7447 & 0.161 & 0.565 \\
14.8982.1963   &05:36:05.0& -71:01:56 & 0.367633 & 0.7448 & 0.120 & 0.633 \\
80.6830.2303   &05:22:15.9& -69:20:43 & 0.367640 & 0.7446 & 0.125 & 0.608 \\
18.2601.439    &04:56:30.3& -68:54:44 & 0.367645 & 0.7444 & 0.124 & 0.927 \\
80.7072.1233   &05:23:47.3& -69:18:19 & 0.367710 & 0.7456 & 0.126 & 0.429 \\
82.8163.1039   &05:30:30.4& -69:12:46 & 0.367774 & 0.7447 & 0.142 & 0.655 \\
5.5136.5295    &05:11:44.3& -69:21:40 & 0.367867 & 0.7442 & 0.105 & 0.705 \\
80.7313.4672   &05:25:34.2& -69:24:16 & 0.368165 & 0.7446 & 0.147 & 0.741 \\
12.9962.1419   &05:41:49.1& -70:16:24 & 0.368761 & 0.7443 & 0.147 & 0.571 \\
13.6804.494    &05:22:21.3& -71:05:07 & 0.369552 & 0.7449 & 0.126 & 0.429 \\
10.4042.1202   &05:05:24.3& -69:40:04 & 0.370325 & 0.7444 & 0.137 & 0.584 \\
80.7321.1365   &05:25:16.5& -68:52:07 & 0.370618 & 0.7442 & 0.131 & 0.817 \\
9.5358.566     &05:13:39.0& -70:41:20 & 0.371015 & 0.7450 & 0.127 & 0.661 \\
6.5850.933     &05:16:15.4& -70:08:14 & 0.373065 & 0.7447 & 0.102 & 0.784 \\
5.5247.894     &05:12:38.0& -69:57:54 & 0.373214 & 0.7447 & 0.118 & 0.805 \\
6.6572.919     &05:21:16.6& -70:22:03 & 0.375691 & 0.7448 & 0.111 & 0.514 \\
6.5721.352     &05:16:05.9& -70:39:36 & 0.376642 & 0.7460 & 0.116 & 0.586 \\
2.4789.1029    &05:09:57.6& -68:17:04 & 0.382408 & 0.7461 & 0.132 & 0.326 \\
5.5369.1184    &05:13:39.0& -69:54:23 & 0.383798 & 0.7457 & 0.129 & 0.605 \\
47.2247.648    &04:54:06.1& -68:19:46 & 0.392076 & 0.7448 & 0.135 & 0.607 \\
10.3557.827    &05:02:18.2& -69:42:31 & 0.393324 & 0.7455 & 0.123 & 0.537 \\
11.9231.545    &05:37:10.7& -70:37:11 & 0.393474 & 0.7461 & 0.089 & 0.472 \\
5.5492.1293    &05:13:57.2& -69:47:06 & 0.394394 & 0.7461 & 0.142 & 0.493 \\
10.4041.840    &05:05:33.2& -69:45:00 & 0.398820 & 0.7461 & 0.117 & 0.504 \\
19.4182.345    &05:06:27.0& -68:24:43 & 0.401625 & 0.7457 & 0.129 & 0.364 \\
10.3191.363    &05:00:33.5& -69:55:49 & 0.406071 & 0.7456 & 0.110 & 0.591 \\
11.8867.861    &05:35:25.9& -70:41:30 & 0.413194 & 0.7465 & 0.128 & 0.312 \\
19.4785.5170   &05:09:47.1& -68:33:13 & 0.428630 & 0.7458 & 0.112 & 0.259 \\
\enddata
\end{deluxetable}
}

\newpage

%
%

\begin{deluxetable}{lccccrc}
\tablecaption{RR1$-\nu1$ variables of the MACHO RR1 sample}
\tablenum{2}
\tablehead {
{\it MACHO \#}& $\alpha$ & $\delta$& $\nu_0$ & $A_0$& $\nu_1-\nu_0$& 
$A_1/A_0$ }
\startdata
47.2609.56      & 04:56:45.1 & -68:21:41 &  2.79858 &  0.112 &  $ -0.01421 $  &  0.268 \\
14.9225.776     & 05:37:19.7 & -70:59:08 &  2.81588 &  0.174 &  $ -0.05414 $  &  0.241 \\
6.6326.424      & 05:19:30.1 & -70:39:32 &  3.02616 &  0.169 &  $ -0.07267 $  &  0.195 \\
6.6091.877      & 05:18:01.8 & -70:11:47 &  3.11905 &  0.062 &  $ -0.03855 $  &  0.694 \\
13.5714.442     & 05:15:29.3 & -71:08:09 &  3.15586 &  0.067 &  \ \ 0.10922 &  0.716 \\
80.6352.1495    & 05:19:37.7 & -68:56:37 &  3.24687 &  0.070 &  $ -0.07175 $  &  0.671 \\
13.6204.617     & 05:18:41.2 & -70:43:53 &  3.41120 &  0.107 &  $ -0.05116 $  &  0.271 \\
15.10072.918    & 05:42:11.3 & -70:58:22 &  3.41207 &  0.195 &  $ -0.03687 $  &  0.210 \\
10.3552.745     & 05:02:37.6 & -70:03:31 &  3.42121 &  0.072 &  \ \ 0.03745 &  0.611 \\
5.5489.1397     & 05:14:36.8 & -69:58:30 &  3.44866 &  0.143 &  $ -0.04786 $  &  0.350 \\
10.4161.1053    & 05:06:14.6 & -69:47:06 &  3.47877 &  0.100 &  \ \ 0.09394 &  0.380 \\
15.11036.255    & 05:48:09.0 & -71:17:19 &  3.48510 &  0.070 &  \ \ 0.07820 &  0.829 \\
11.9471.780     & 05:38:33.6 & -70:43:58 &  3.49672 &  0.160 &  $ -0.11346 $  &  0.275 \\
9.5242.1032     & 05:12:43.9 & -70:20:02 &  3.49779 &  0.126 &  \ \ 0.03749 &  0.563 \\
13.5713.590     & 05:15:25.7 & -71:10:31 &  3.52566 &  0.076 &  \ \ 0.10042 &  0.553 \\
82.8289.887     & 05:31:19.2 & -68:50:33 &  3.53404 &  0.103 &  $ -0.18970 $  &  0.311 \\
2.5266.3864     & 05:12:36.8 & -68:43:38 &  3.57930 &  0.127 &  \ \ 0.06554 &  0.370 \\
80.7072.2280    & 05:23:45.4 & -69:20:54 &  3.58891 &  0.105 &  $ -0.04727 $  &  0.257 \\
80.7437.1678    & 05:26:25.3 & -69:12:11 &  3.59531 &  0.128 &  $ -0.08894 $  &  0.688 \\
6.5729.958      & 05:16:01.3 & -70:06:44 &  3.59948 &  0.117 &  \ \ 0.07423 &  0.761 \\
3.6603.795      & 05:20:41.0 & -68:20:21 &  3.61654 &  0.179 &  $ -0.01735 $  &  0.218 \\
6.5730.4057     & 05:16:02.0 & -70:05:10 &  3.61898 &  0.077 &  \ \ 0.07758 &  0.636 \\
14.9702.401     & 05:40:17.6 & -71:26:48 &  3.63103 &  0.163 &  $ -0.18408 $  &  0.196 \\
82.8766.1305    & 05:34:23.3 & -69:20:04 &  3.85726 &  0.080 &  $ -0.00306 $  &  0.550 \\
\enddata
\tablecomments{\\ 
13.6204.617  -- `b' data support RR1 status \\
47.2609.56   -- also an RR01 variable with $P_1/P_0 = 0.7443$ 
and $A_1/A_0 = 0.22$ \\
6.5729.958   -- in `b' there is a third component at 
$P(\nu_2)/P(\nu_0) = 0.757$ \\
82.8766.1305 -- hint for RR1$-BL$ classification in the `b' data }
\end{deluxetable}
%
%

%
%

\newpage

{\footnotesize
\begin{deluxetable}{lcccccc}
\tablecaption{RR1$-BL$ variables of the MACHO RR1 sample}
\tablenum{3}
\tablehead{
{\it MACHO \#}& $\alpha$ & $\delta$& $f_0$ & $A_0$& $\Delta f_-$ & $A_-/A_0$ \\ &  &  &  & & $\Delta f_+$ & $A_+/A_0$ }
\startdata
81.8758.1447    & 05:34:39.1 & -69:51:33 &  2.19180 &  0.283 &  $ -0.02987 $  &  0.117 \\ &  &  &  &  &  \ \ 0.02991 &  0.170 \\
6.7054.713      & 05:24:03.1 & -70:33:38 &  2.27063 &  0.152 &  $ -0.00166 $  &  0.224 \\ &  &  &  &  &  \ \ 0.00163 &  0.132 \\
2.5271.255      & 05:13:05.0 & -68:22:49 &  2.30017 &  0.113 &  $ -0.00129 $  &  0.345 \\ &  &  &  &  &  \ \ 0.00121 &  0.469 \\
80.6957.409     & 05:23:27.6 & -68:55:38 &  2.45168 &  0.104 &  $ -0.00084 $  &  0.183 \\ &  &  &  &  &  \ \ 0.00079 &  0.298 \\
3.6240.450      & 05:19:03.1 & -68:20:21 &  2.60663 &  0.155 &  $ -0.00099 $  &  0.258 \\ &  &  &  &  &  \ \ 0.00104 &  0.271 \\
81.8639.1749    & 05:33:52.6 & -69:43:21 &  2.66022 &  0.135 &  $ -0.04469 $  &  0.200 \\ &  &  &  &  &  \ \ 0.04482 &  0.252 \\
19.4188.1264    & 05:06:01.5 & -68:00:45 &  2.70082 &  0.127 &  $ -0.00096 $  &  0.402 \\ &  &  &  &  &  \ \ 0.00105 &  0.591 \\
80.6953.1751    & 05:23:31.4 & -69:11:07 &  2.84322 &  0.182 &  $ -0.02811 $  &  0.280 \\ &  &  &  &  &  \ \ 0.02808 &  0.341 \\
15.10311.782    & 05:44:00.5 & -71:10:42 &  2.87117 &  0.146 &  $ -0.00106 $  &  0.240 \\ &  &  &  &  &  \ \ 0.00118 &  0.226 \\
5.5368.1201     & 05:13:44.2 & -70:00:45 &  2.97166 &  0.197 &  $ -0.00558 $  &  0.264 \\ &  &  &  &  &  \ \ 0.00552 &  0.137 \\
80.6595.1599    & 05:21:08.3 & -68:51:17 &  3.00869 &  0.128 &  $ -0.03684 $  &  0.484 \\ &  &  &  &  &  \ \ 0.03683 &  0.250 \\
18.2361.870     & 04:54:43.2 & -68:48:06 &  3.07240 &  0.137 &  $ -0.02973 $  &  0.263 \\ &  &  &  &  &  \ \ 0.02974 &  0.358 \\
82.8526.1176    & 05:32:47.6 & -69:11:52 &  3.08750 &  0.066 &  $ -0.12705 $  &  0.288 \\ &  &  &  &  &  \ \ 0.12704 &  0.227 \\
9.5479.852      & 05:14:08.3 & -70:39:32 &  3.10820 &  0.154 &  $ -0.00069 $  &  0.532 \\ &  &  &  &  &  \ \ 0.00078 &  0.675 \\
14.9223.737     & 05:37:36.3 & -71:06:35 &  3.22706 &  0.148 &  $ -0.03161 $  &  0.439 \\ &  &  &  &  &  \ \ 0.03159 &  0.486 \\
82.8765.1250    & 05:34:02.0 & -69:21:56 &  3.27951 &  0.159 &  $ -0.04898 $  &  0.252 \\ &  &  &  &  &  \ \ 0.04900 &  0.264 \\
82.8049.746     & 05:29:34.8 & -68:45:34 &  3.34744 &  0.162 &  $ -0.07194 $  &  0.278 \\ &  &  &  &  &  \ \ 0.07191 &  0.247 \\
15.10068.239    & 05:42:11.3 & -71:14:50 &  3.35914 &  0.124 &  $ -0.03338 $  &  0.274 \\ &  &  &  &  &  \ \ 0.03351 &  0.177 \\
82.8408.1002    & 05:32:00.7 & -68:58:50 &  3.37504 &  0.163 &  $ -0.03544 $  &  0.221 \\ &  &  &  &  &  \ \ 0.03542 &  0.325 \\
2.5148.1207     & 05:12:23.9 & -68:29:43 &  3.39346 &  0.121 &  $ -0.03528 $  &  0.289 \\ &  &  &  &  &  \ \ 0.03526 &  0.273 \\
15.10313.606    & 05:43:45.0 & -71:04:54 &  3.41874 &  0.128 &  $ -0.06431 $  &  0.422 \\ &  &  &  &  &  \ \ 0.06438 &  0.477 \\
13.6810.2992    & 05:22:45.6 & -70:39:11 &  3.44356 &  0.125 &  $ -0.04147 $  &  0.664 \\ &  &  &  &  &  \ \ 0.04159 &  0.456 \\
6.5971.1233     & 05:16:59.0 & -70:08:14 &  3.47563 &  0.120 &  $ -0.07350 $  &  0.350 \\ &  &  &  &  &  \ \ 0.07351 &  0.175 \\
19.4188.195     & 05:06:26.3 & -67:58:57 &  3.53583 &  0.123 &  $ -0.01215 $  &  0.358 \\ &  &  &  &  &  \ \ 0.01210 &  0.431 \\
80.7441.933     & 05:26:06.4 & -68:56:02 &  3.65881 &  0.130 &  $ -0.04795 $  &  0.246 \\ &  &  &  &  &  \ \ 0.04800 &  0.338 \\
11.9355.1380    & 05:38:21.3 & -70:23:22 &  3.70248 &  0.039 &  $ -0.00465 $  &  2.308 \\ &  &  &  &  &  \ \ 0.00466 &  1.333 \\
5.4401.1018     & 05:07:33.5 & -69:55:12 &  3.72395 &  0.076 &  $ -0.00480 $  &  0.816 \\ &  &  &  &  &  \ \ 0.00470 &  0.579 \\
9.5125.1018     & 05:11:47.0 & -70:05:37 &  3.84901 &  0.108 &  $ -0.00520 $  &  0.500 \\ &  &  &  &  &  \ \ 0.00519 &  0.204 \\
\enddata
\tablecomments{\\ 
5.5368.1201 -- `b' data are used \\
9.5479.852  -- `b' data are used }
\end{deluxetable}
}

%
%

\newpage

{\footnotesize
\begin{deluxetable}{lcccclccc}
\tablecaption{RR1$-PC$ variables of the MACHO RR1 sample}
\tablenum{4}
\tablehead {
{\it MACHO \#}& $\alpha$ & $\delta$& $P_1$ &  &{\it MACHO \#}& $\alpha$ & $\delta$& $P_1$ }
\startdata
2.5389.1478     & 05:13:40.9 & -68:37:18 & 0.250558 &  & 3.6362.689      & 05:19:32.1 & -68:16:48 & 0.357375 \\
3.7450.214      & 05:26:25.7 & -68:19:52 & 0.281339 &  & 18.2597.765     & 04:56:08.5 & -69:12:11 & 0.357969 \\
6.6094.5406     & 05:18:07.1 & -70:00:05 & 0.285164 &  & 80.6838.2884    & 05:22:18.3 & -68:46:38 & 0.358339 \\
6.6810.616      & 05:22:16.7 & -70:39:50 & 0.288385 &  & 80.7072.1545    & 05:23:45.0 & -69:20:51 & 0.358344 \\
14.8497.534     & 05:32:39.9 & -71:07:44 & 0.289300 &  & 9.4873.497      & 05:10:45.8 & -70:44:40 & 0.358621 \\
13.6326.1733    & 05:19:49.9 & -70:38:49 & 0.290353 &  & 6.5848.1122     & 05:16:14.7 & -70:17:33 & 0.359143 \\
13.6077.638     & 05:18:11.5 & -71:08:05 & 0.290690 &  & 14.9590.3902    & 05:39:26.8 & -70:50:32 & 0.360039 \\
9.4394.386      & 05:07:58.5 & -70:21:54 & 0.290778 &  & 5.4766.918      & 05:10:02.5 & -69:46:33 & 0.361072 \\
10.3557.1024    & 05:02:42.7 & -69:43:47 & 0.294767 &  & 2.5872.1272     & 05:16:21.1 & -68:37:57 & 0.362671 \\
10.3556.986     & 05:02:04.4 & -69:48:28 & 0.296149 &  & 9.5363.899      & 05:13:34.6 & -70:21:08 & 0.364048 \\
81.8879.1869    & 05:34:52.9 & -69:51:52 & 0.297565 &  & 12.10801.843    & 05:47:04.2 & -70:49:30 & 0.365089 \\
80.7319.1287    & 05:25:31.4 & -69:01:20 & 0.297860 &  & 13.6198.531     & 05:18:47.6 & -71:05:58 & 0.366172 \\
9.5599.762      & 05:15:17.5 & -70:44:55 & 0.299416 &  & 9.4994.491      & 05:11:22.0 & -70:44:24 & 0.366418 \\
9.4760.861      & 05:09:36.4 & -70:13:23 & 0.299667 &  & 11.9113.1731    & 05:36:38.0 & -70:23:21 & 0.366441 \\
12.10808.830    & 05:47:01.6 & -70:18:09 & 0.300015 &  & 15.10914.663    & 05:47:22.6 & -71:20:57 & 0.367050 \\
9.4761.1258     & 05:09:58.6 & -70:08:13 & 0.300939 &  & 10.4040.917     & 05:05:14.1 & -69:47:53 & 0.367383 \\
11.9116.1382    & 05:36:19.9 & -70:13:36 & 0.302487 &  & 3.6598.939      & 05:21:03.0 & -68:38:22 & 0.367820 \\
80.7436.1463    & 05:26:04.6 & -69:16:12 & 0.302509 &  & 3.6723.769      & 05:21:22.8 & -68:23:46 & 0.369235 \\
9.4392.833      & 05:07:35.7 & -70:30:51 & 0.303440 &  & 82.8283.1040    & 05:31:19.5 & -69:17:37 & 0.370077 \\
18.3086.1308    & 04:59:08.1 & -68:53:28 & 0.304942 &  & 13.5844.3963    & 05:16:10.1 & -70:32:08 & 0.370272 \\
15.10795.916    & 05:47:00.0 & -71:11:37 & 0.307457 &  & 6.6212.1142     & 05:19:04.2 & -70:11:59 & 0.373232 \\
5.4767.1388     & 05:09:50.0 & -69:44:26 & 0.307806 &  & 82.8406.1180    & 05:31:50.3 & -69:08:37 & 0.373883 \\
80.6597.4435    & 05:20:39.1 & -68:45:14 & 0.308645 &  & 82.8288.869     & 05:31:02.9 & -68:55:13 & 0.374056 \\
10.3797.865     & 05:03:46.3 & -69:53:20 & 0.308792 &  & 5.4524.972      & 05:08:42.3 & -69:45:56 & 0.374232 \\
80.7195.1166    & 05:25:01.1 & -69:11:08 & 0.310755 &  & 13.6080.541     & 05:18:10.1 & -70:57:30 & 0.374479 \\
80.6951.2395    & 05:23:23.6 & -69:18:38 & 0.311420 &  & 15.10069.680    & 05:42:26.9 & -71:13:33 & 0.374508 \\
19.4669.2905    & 05:09:21.0 & -68:10:27 & 0.311959 &  & 2.4908.826      & 05:10:57.3 & -68:22:22 & 0.376579 \\
2.5630.1156     & 05:15:02.5 & -68:39:15 & 0.312314 &  & 5.5008.1902     & 05:11:22.8 & -69:48:30 & 0.377428 \\
2.5151.924      & 05:12:06.2 & -68:18:42 & 0.314624 &  & 19.4429.678     & 05:07:36.8 & -68:02:37 & 0.377518 \\
2.5514.660      & 05:14:27.9 & -68:18:54 & 0.315373 &  & 5.4769.1363     & 05:10:01.1 & -69:34:38 & 0.378528 \\
3.7444.782      & 05:26:23.2 & -68:44:13 & 0.317451 &  & 11.9355.1235    & 05:38:25.6 & -70:23:42 & 0.379060 \\
81.9004.1177    & 05:35:57.0 & -69:33:52 & 0.318030 &  & 12.11042.810    & 05:48:12.8 & -70:51:37 & 0.379823 \\
9.4875.697      & 05:10:45.0 & -70:35:11 & 0.320646 &  & 3.7205.727      & 05:24:30.4 & -68:31:31 & 0.381238 \\
2.5148.713      & 05:11:42.8 & -68:32:12 & 0.321252 &  & 6.6813.699      & 05:22:13.9 & -70:28:34 & 0.382572 \\
5.4891.1617     & 05:10:26.3 & -69:33:27 & 0.321780 &  & 6.5724.620      & 05:15:58.0 & -70:28:48 & 0.382936 \\
13.5721.2180    & 05:16:02.7 & -70:39:39 & 0.322156 &  & 81.9123.806     & 05:36:40.7 & -69:43:09 & 0.385314 \\
11.9594.880     & 05:39:42.8 & -70:37:17 & 0.322277 &  & 12.10924.961    & 05:47:58.7 & -70:38:07 & 0.386863 \\
6.6933.1043     & 05:23:22.3 & -70:32:35 & 0.322842 &  & 2.5388.1150     & 05:13:35.5 & -68:39:42 & 0.388582 \\
3.7088.623      & 05:24:15.8 & -68:16:34 & 0.324418 &  & 81.9723.894     & 05:40:15.9 & -70:02:56 & 0.388602 \\
82.8282.1019    & 05:31:08.0 & -69:18:29 & 0.327110 &  & 5.5489.1125     & 05:14:33.3 & -70:01:07 & 0.389857 \\
80.7436.1633    & 05:26:10.9 & -69:14:48 & 0.330471 &  & 80.6350.3508    & 05:19:37.6 & -69:05:23 & 0.390063 \\
9.5600.566      & 05:15:04.5 & -70:41:39 & 0.331331 &  & 2.5876.444      & 05:16:27.5 & -68:25:14 & 0.396680 \\
5.4766.1109     & 05:09:48.8 & -69:48:47 & 0.332465 &  & 6.6697.1361     & 05:21:35.2 & -70:09:35 & 0.397333 \\
6.6452.2394     & 05:20:25.3 & -70:17:55 & 0.334832 &  & 9.4634.863      & 05:08:45.1 & -70:33:10 & 0.398111 \\
3.6240.470      & 05:19:06.4 & -68:20:54 & 0.334995 &  & 9.4879.502      & 05:10:49.5 & -70:18:28 & 0.398205 \\
6.6455.1326     & 05:20:23.8 & -70:09:14 & 0.335632 &  & 9.5364.886      & 05:13:45.4 & -70:16:26 & 0.398766 \\
6.6094.5331     & 05:18:17.8 & -70:00:14 & 0.335685 &  & 13.6080.628     & 05:18:18.5 & -70:55:58 & 0.400474 \\
2.4787.770      & 05:10:11.9 & -68:22:56 & 0.335706 &  & 6.5971.1194     & 05:17:36.9 & -70:08:26 & 0.400978 \\
81.9723.796     & 05:40:19.4 & -70:05:40 & 0.336952 &  & 6.6451.919      & 05:20:22.7 & -70:24:56 & 0.402077 \\
19.3823.546     & 05:03:43.2 & -68:06:29 & 0.337930 &  & 13.6441.529     & 05:20:19.2 & -71:05:14 & 0.402531 \\
13.6568.3005    & 05:20:42.8 & -70:38:46 & 0.338099 &  & 11.9349.558     & 05:37:49.0 & -70:47:36 & 0.402859 \\
80.6950.6414    & 05:23:35.5 & -69:21:54 & 0.338445 &  & 13.6080.594     & 05:18:18.0 & -70:56:08 & 0.404710 \\
6.6094.5606     & 05:18:13.2 & -69:58:41 & 0.339385 &  & 81.8875.2038    & 05:34:57.0 & -70:05:56 & 0.405794 \\
15.10433.787    & 05:44:52.1 & -71:06:48 & 0.339489 &  & 80.7192.3592    & 05:25:02.3 & -69:24:50 & 0.405959 \\
80.7440.1192    & 05:26:28.6 & -68:59:17 & 0.343025 &  & 80.6953.1590    & 05:23:23.6 & -69:10:17 & 0.406129 \\
14.9346.412     & 05:38:13.8 & -70:58:07 & 0.343495 &  & 82.8041.1029    & 05:30:00.7 & -69:16:53 & 0.406715 \\
13.6441.527     & 05:20:08.9 & -71:05:36 & 0.343520 &  & 19.3823.473     & 05:03:42.4 & -68:07:30 & 0.408980 \\
80.6468.2765    & 05:20:30.5 & -69:13:45 & 0.349369 &  & 81.8518.1621    & 05:33:02.2 & -69:42:37 & 0.410623 \\
6.5729.1008     & 05:15:47.5 & -70:06:08 & 0.350106 &  & 9.4390.560      & 05:07:17.6 & -70:38:15 & 0.411036 \\
80.6835.1220    & 05:22:30.0 & -68:57:42 & 0.350229 &  & 9.5360.903      & 05:13:29.3 & -70:31:35 & 0.413897 \\
10.4165.348     & 05:06:01.3 & -69:32:29 & 0.350558 &  & 11.8744.658     & 05:34:10.4 & -70:48:10 & 0.415590 \\
19.4784.5754    & 05:09:51.5 & -68:34:28 & 0.351798 &  & 14.8744.3716    & 05:34:10.6 & -70:48:13 & 0.415694 \\
2.5870.4598     & 05:16:38.9 & -68:49:28 & 0.352372 &  & 11.9235.1034    & 05:37:13.5 & -70:18:43 & 0.417598 \\
15.10308.620    & 05:43:55.6 & -71:23:39 & 0.353552 &  & 2.5273.1267     & 05:12:28.9 & -68:16:07 & 0.420892 \\
81.9724.295     & 05:40:22.2 & -69:59:22 & 0.353864 &  & 13.6076.306     & 05:18:15.4 & -71:13:35 & 0.427092 \\
11.9471.1050    & 05:38:43.5 & -70:42:47 & 0.354204 &  & 3.6236.690      & 05:18:36.8 & -68:36:38 & 0.435675 \\
3.6360.656      & 05:19:11.9 & -68:23:37 & 0.354548 &  & 13.6926.490     & 05:22:56.2 & -71:00:12 & 0.438382 \\
6.6812.923      & 05:22:26.5 & -70:30:19 & 0.354785 &  & 2.5271.1540     & 05:12:27.7 & -68:25:31 & 0.439438 \\
5.5250.1501     & 05:13:09.0 & -69:46:38 & 0.355176 &  & 82.8890.449     & 05:34:44.2 & -69:08:29 & 0.440049 \\
2.4663.944      & 05:09:04.9 & -68:34:50 & 0.355264 &  & 11.8867.970     & 05:35:14.5 & -70:39:11 & 0.446641 \\
6.6692.937      & 05:21:29.1 & -70:29:23 & 0.356880 \\
\enddata
\end{deluxetable}
}

%
%

\newpage

\begin{deluxetable}{lcccl}
\tablecaption{Miscellaneous variables of the MACHO RR1 sample}
\tablenum{5}
\tablehead{ {\it MACHO \#} & $\alpha$ & $\delta$& $P_1$ & {\it Type} }
\startdata
9.4278.179      & 05:07:03.5 & -70:04:06 & 0.326816 & RR12        \\
12.10443.367    & 05:44:36.9 & -70:28:50 & 0.336559 & RR12        \\
12.10202.285    & 05:43:04.3 & -70:22:47 & 0.398114 & RR12        \\
15.9947.338     & 05:41:43.0 & -71:14:05 & 0.288221 & RR1$-\nu2$  \\
11.8987.787     & 05:36:05.9 & -70:44:43 & 0.304127 & RR1$-\nu2$  \\
2.5023.5787     & 05:11:27.9 & -68:47:50 & 0.324536 & RR1$-\nu2$  \\
14.8376.548     & 05:32:15.3 & -71:08:10 & 0.291571 & RR1$-\nu M$ \\
11.8751.1740    & 05:34:39.9 & -70:20:55 & 0.357751 & RR1$-\nu M$ \\
19.4671.684     & 05:09:14.5 & -68:05:30 & 0.378744 & RR1$-\nu M$ \\
5.5128.1262     & 05:11:43.2 & -69:50:57 & 0.427975 & RR1$-\nu M$ \\
6.6697.1565     & 05:21:49.6 & -70:06:58 & 0.433493 & RR1$-\nu M$ \\
12.11043.1000   & 05:48:19.1 & -70:46:08 & 0.264320 & RR1$-NC$    \\
9.5123.633      & 05:11:45.6 & -70:12:17 & 0.267730 & RR1$-NC$    \\
82.8525.1980    & 05:32:52.3 & -69:13:51 & 0.283398 & RR1$-NC$    \\
9.5239.1141     & 05:12:44.3 & -70:31:21 & 0.363336 & RR1$-NC$    \\
13.6802.544     & 05:22:43.5 & -71:10:48 & 0.378060 & RR1$-NC$    \\
10.4403.4871    & 05:07:21.4 & -69:46:46 & 0.457304 & RR1$-NC$    \\
6.6093.5030     & 05:18:05.1 & -70:03:56 & 0.265360 & RR1$-D1$    \\
3.6243.404      & 05:18:42.4 & -68:06:50 & 0.270851 & RR1$-D1$    \\
80.6353.1458    & 05:19:36.8 & -68:49:41 & 0.275324 & RR1$-D1$    \\
9.5599.617      & 05:14:52.0 & -70:45:24 & 0.278206 & RR1$-D1$    \\
80.6708.6771    & 05:21:27.0 & -69:23:22 & 0.286094 & RR1$-D1$    \\
6.6815.747      & 05:22:31.1 & -70:21:39 & 0.292024 & RR1$-D1$    \\
9.5004.750      & 05:11:36.8 & -70:04:57 & 0.304161 & RR1$-D1$    \\
81.8521.1454    & 05:33:13.2 & -69:31:35 & 0.309440 & RR1$-D1$    \\
3.6961.824      & 05:22:55.7 & -68:39:54 & 0.309894 & RR1$-D1$    \\
9.5122.363      & 05:11:47.5 & -70:16:17 & 0.322232 & RR1$-D1$    \\
6.6931.649      & 05:23:01.5 & -70:39:45 & 0.322806 & RR1$-D1$    \\
12.10920.615    & 05:47:43.4 & -70:54:12 & 0.323981 & RR1$-D1$    \\
5.4649.1029     & 05:09:08.2 & -69:30:52 & 0.325266 & RR1$-D1$    \\
13.6806.664     & 05:22:17.6 & -70:54:28 & 0.330759 & RR1$-D1$    \\
80.6709.2322    & 05:21:37.9 & -69:21:26 & 0.332538 & RR1$-D1$    \\
9.5360.768      & 05:13:34.5 & -70:29:57 & 0.337518 & RR1$-D1$    \\
14.8854.199     & 05:34:55.9 & -71:29:44 & 0.339066 & RR1$-D1$    \\
81.8519.1395    & 05:32:48.6 & -69:41:06 & 0.339398 & RR1$-D1$    \\
12.11283.284    & 05:49:38.5 & -70:55:33 & 0.340384 & RR1$-D1$    \\
6.6699.5598     & 05:21:34.6 & -70:01:18 & 0.352686 & RR1$-D1$    \\
\enddata
\end{deluxetable}

%
%

\begin{deluxetable}{lccccccl}
\tablecaption{RR12 and RR1$-\nu2$ variables of the MACHO RR1 sample}
\tablenum{6}
\tablehead{
{\it MACHO \#}& $\nu_0$ & $\nu_1$ & $\nu_2$ & $\nu_0/\nu_1$& 
$A_1/A_0$ & $A_2/A_0$ & {\it Type} }
\startdata
12.10202.285 & 2.511844 & 3.118910 & --- & 0.8054 & 0.260 & --- & RR12\\
12.10443.367 & 2.971251 & 3.701164 & --- & 0.8028 & 0.294 & --- & RR12\\
9.4278.179   & 3.059812 & 3.801934 & --- & 0.8048 & 0.296 & --- & RR12\\
2.5023.5787  & 3.081326 & 3.068358 & 3.073338 & 1.0042 & 1.000 & 0.926 
& RR1$-\nu2$\\
11.8987.787  & 3.288101 & 3.238294 & 3.129587 & 1.0154 & 0.638 & 0.345 
& RR1$-\nu2$\\
15.9947.338  & 3.469549 & 3.421443 & 3.434602 & 1.0141 & 0.559 & 0.419 
& RR1$-\nu2$\\
\enddata
\end{deluxetable}

\end{document}